\documentclass[twocolumn,amssymb, nobibnotes, floatfix, aps, prb]{revtex4-2}
\usepackage{amssymb}
\usepackage{amsmath}
\usepackage{graphicx}
\usepackage{color}
\usepackage{physics}
\usepackage{mathtools}
\renewcommand{\vec}[1]{\boldsymbol{#1}}
\begin{document}

%\preprint{}

\title{Continuum models for twisted bilayer graphene: the effects of lattice deformation and hopping parameters}
%\title{Lattice Deformation, Low Energy Models and Flat Bands in Twisted Bilayer Graphene}

\author{Francisco Guinea$^{1,2}$}
\email{Francisco.Guinea@imdea.org}
\author{Niels R. Walet$^1$}
\email{Niels.Walet@manchester.ac.uk}
\homepage{https://www.research.manchester.ac.uk/portal/niels.walet.html}
%\thanks{}
\affiliation{$^1$School of Physics and Astronomy, University of Manchester, Manchester, M13 9PY, UK}
\affiliation{$^2$Imdea Nanoscience, Faraday 9, 28015 Madrid, Spain}

\date{\today}

\begin{abstract}
We analyze a description of twisted graphene bilayers, that incorporates the deformation of the layers using state of the art interlayer atomic potentials, and a modification of the hopping parameters between layers in the light of the classic Slonczewski-Weiss-McClure parametrisation.
We obtain narrow bands in all cases, but that their nature can be rather different. We will show
how to describe the results by equivalent continuum models. Even though such models can be constructed, their complexity can vary, requiring many coupling parameters to be included, and the full in-layer dispersion must be taken into account.
The combination of all these effects will  have a large impact on the wave functions of the flat bands, and that modifications in details of the underlying models can lead to significant changes. A robust conclusion is that the natural strength of the interlayer couplings is higher than usually assumed, leading to shifts in the definition of the magic angles. The structure at the edges of the narrow bands, at the $\Gamma$ point of the Brillouin Zone is also strongly dependent on parametrization. As a result, the existence, and size, of band gaps between the flat bands and the neighboring ones are changed. Hence, the definition of Wannier functions, and descriptions based on local interactions are strongly dependent on the description of the model at the atomic scale.
\end{abstract}

\pacs{???}
%\keywords{}

\maketitle
\section{Introduction}
The discovery of strong interactions and superconductivity in twisted graphene bilayers has been one of the main achievement in two-dimensional materials in the past year; it has been chosen as the Physics World breakthrough of the year 2018 \cite{Ketal17,Cetal18a,Cetal18b}, see also Ref.~\cite{Hetal18}. This field has grown so rapidly that it now carries its own dedicated label, ``twistronics". Twisted graphene layers show a rich phenomenology, likely due to the interplay of a complex electronic structure and the effects of electron interactions. The core ideas build on previous work on the behavior of graphene superlattices on a BN substrate, see for example Refs.~\cite{Letal11,Letal11b,Yetal11,Petal13,Hetal13,Detal13}. In all of these cases we have a periodic, long wavelength, Moir\'e modulation, but for graphene on BN  the mismatch in lattice spacing between the different materials in the layers limits the maximum wavelength, and thus the diversity of electronic structures for the accessible modulations \cite{WPMGF13,JRQM14,SGSG14}.  On the other hand, the two graphene layers in a twisted bilayer have the same spacing and the periodicity of the Moir\'e structure has no limit, and diverges at small twist angles \cite{LPN07,M10,SCVPB10,bistritzer_moire_2011,M11,LPN12}, $L_M = d / \bigl(2 \sin ( \theta / 2 )\bigr)$, where $d \approx 2.42\,\text{\AA}$ is the lattice unit of graphene. For sufficiently small angles almost flat bands arise near the charge neutrality point \cite{TMM10,SCVPB10,bistritzer_moire_2011}. The effects of the intrinsically small interaction effects in graphene are expected to be enhanced for special `magic' angles where the width of the low energy bands is smallest. Novel magnetic phases become possible when the lowest band is half filled \cite{GLGS17}. Layer dependent strains can also lead to Moir\'e structures and narrow bands \cite{SGG12,Hetal18b}.

\begin{figure}
\includegraphics[width=6cm]{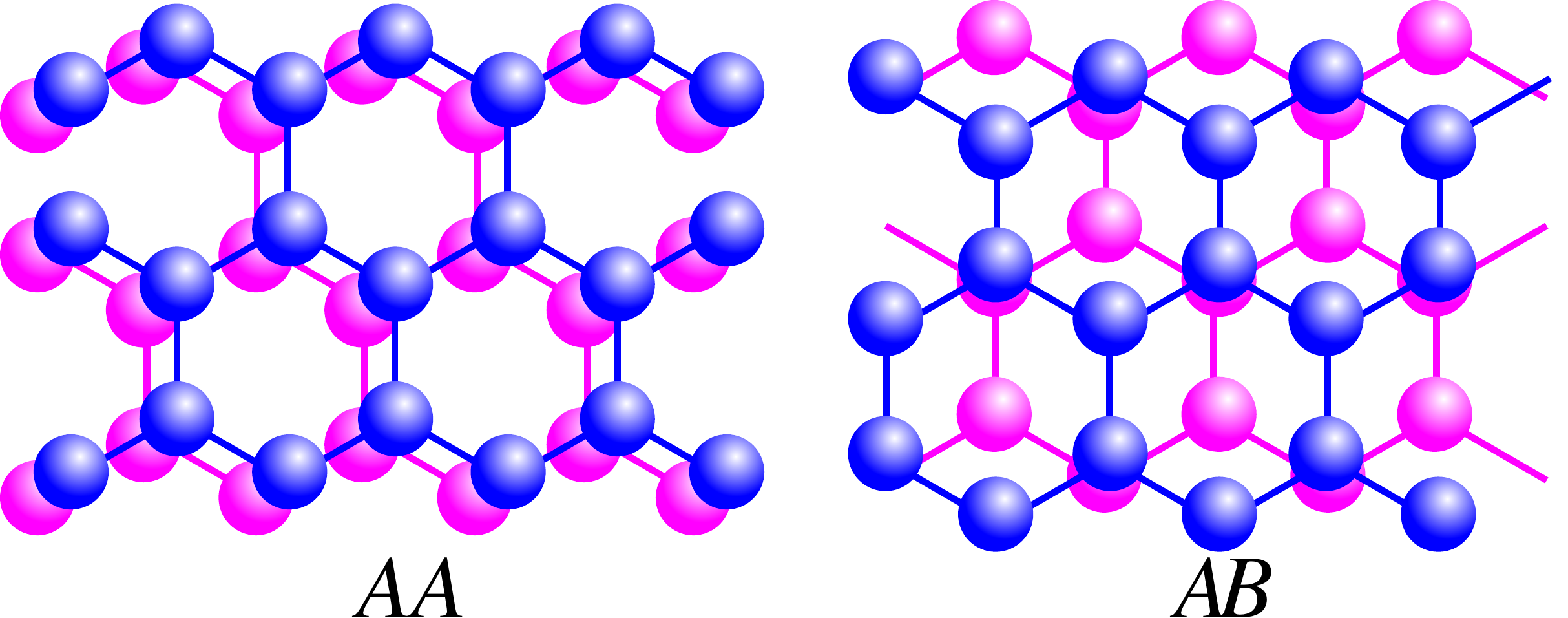}
\caption{Examples of a graphene bilayer in (approximate) $AA$ and $AB$  alignment.}\label{fig:AAAB}
\end{figure}

When we (almost) align two graphene layers, we have two minimum energy options as shown in Fig.~\ref{fig:AAAB}. We can either replicate the two layers with only a change in the height ($AA$ alignment), or we can translate one of the layers over a single nearest neigbor distance, which gives $AB$ alignment. In areas with $AB$ alignment  half of the carbon atoms in one layer align with those of the other one, but the other half aligns with the midpoints of the hexagons in the other layer. This situation has a lower energy than that with $AA$ alignment. If we consider a twisted bilayer, where both layers are perfectly hexagonal but rotated by an angle relative to a common axis, we find areas with both alignments that are of the same size. At a small cost, the graphene layers can warp, both in and out of plane, to enlarge the beneficial effect of the $AB$ alignment. Doing a fully microscopic calculation (which in this case would require a computationally extremely expensive Green's function Monte Carlo analysis, since density functional theory calculations struggle to describe bilayer graphene \cite{mostaani_quantum_2015}, see also\cite{SWSLFB18,Letal19}) is out of the question for the more than $10,000$ carbon atoms that are contained in a single unit cell, so we need to fall back to simpler  models. A few DFT studies are available in the literature\cite{CMFCLK17,SWSLFB18,Letal19}, although it does not (yet) seem feasible to carry out calculations at the size required to deal with small twist angles.

We can use elegant and simple continuum models when we have no deformation \cite{LPN07,bistritzer_moire_2011}, or we can use semi-microscopic atomistic models, such as classical force models for the interatomic forces, both within each layer and between different layers, combined with tight-binding methods for the electronic structure. 
As we shall discuss below, this latter approach, which relies implicitly on many-body interactions, is likely to give the most realistic description.

\begin{figure}
\includegraphics[width=\columnwidth]{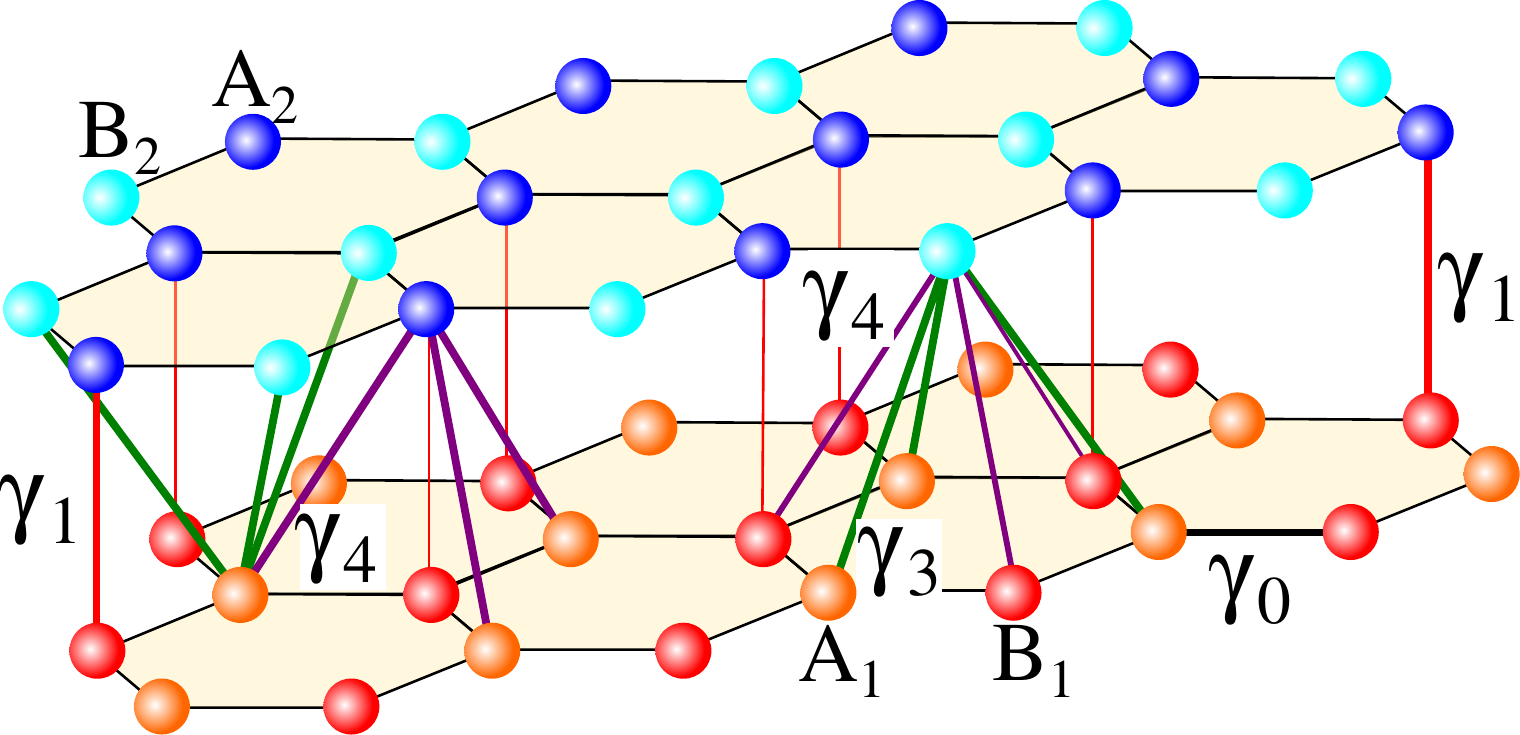}
\caption{The  definition of the hopping parameters $\gamma_i$ as used in the SWM model for $B_1A_2$ aligned layers. We denote $\gamma_0$ by a black line, $\gamma_1$ by a red line, $\gamma_3$ in green and $\gamma_4$ in purple.}\label{fig:SWM}
\end{figure}

At the same time we need to ask ourselves what is the ``best" tight-binding description for the electronic structure: For a single layer of graphene the standard approach is to use a nearest-neighbor hopping, and maybe a next nearest neighbor one, to describe the spectra. That approach  work very well, even for systems with deformed lattices (typically Moir\'e supercells). The structure of classical potential models that describe the atomic positions of the atoms in a 2D layer is well understood, and most modern potential models describe the structure of graphene both near and far from equilibrium very well.

The description of both the binding of a bilayer, and the electronic hopping between the layers is much more challenging. The most realistic potential models contain complex many-body interactions, that are necessary to describe the complexities of intra- and inter-layer binding. 
It is also reasonably well established that one must include many-body effects in the hopping parameters for both graphite and graphene. The key signature of the problems with a two-body description is the difference between nearest-neighbor hopping  parameters for different positions in an $AB$-aligned the lattice. As originally described for graphite in the Slonczewski-Weiss-McClure (SWM) model \cite{slonczewski_band_1958,mcclure_band_1957,mcclure_theory_1960} the hopping parameter $\gamma_1$, between vertically displaced carbon atoms in $B_1 A_2$ alignment, which has a value of about $0.4\,\text{eV}$ in graphene \cite{Brandt88,DD2002,Neto2009}, differs strongly from the two hopping  parameters for slightly larger distances: $\gamma_4=0.04-0.15\,\text{eV}$ for the hopping near vertical alignment ($B_1B_2$, etc.), and $\gamma_3=0.3\,\text{eV}$ for midpoint aligned carbon atoms (usually labelled $A_1B_2$). In Fig.~\ref{fig:SWM} we show how $\gamma_4$ ($B_1B_2$) occurs next to $\gamma_3$, and that both have the same hopping distanc. Nevertheless $\gamma_4$ is much smaller than $\gamma_3$ in graphite, which is not captured by the standard distance-dependent two-center Koster-Slater hopping.  As discussed in a recent review \cite{mccann_electronic_2013},  for bilayer graphene there is a spread in the values found and used. The consensus is that the  value of $\gamma_4$  is still substantially smaller than
$\gamma_3$, see also Ref.~\cite{M11}. 
A useful form of a model where the screening is dominated by in-layer nearest-neighbor atoms is given in
Ref.~\citep{sboychakov_electronic_2015}, see also Refs.~\cite{tang_environment_dependent_1996,SRRN17}.
This is very similar to the case of the interatomic potentials, which also require a many-body screening largely dominated by nearest-neighbors. 
Clearly both the graphene lattice deformation and the many-body effects in the hopping will play an important role in describing the band structure obtained in a tight-binding model. 

Once we have determined the atomic positions and the hopping parameters for the tight-binding model, we need to deal with the large dimensionality which arises from the size of the unit cell, which leads to a large number of bands. Especially for small angles and thus long Moir\'e wavelengths, the matrices become extremely large. However, these matrices are very sparse and can be dealt with sparse matrix methods such as ARPACK \cite{lehoucq1998arpack}. Even using those methods numerical calculations are still time consuming. Thus, especially if we want to study many-body physics, we would like to reduce the full tight-binding model to a more efficient low-energy effective model. The one usually used is  discussed in Refs.~\cite{LPN07,bistritzer_moire_2011}, but only works for the simplest lattice and hopping parameters. Since we will use a more complex tight-binding model than normally considered and lattice deformation on top of that, we need to more be careful in making this reduction. We shall investigate this in detail, using an approach that incorporates and generalises the ideas of Ref.~\citep{koshino_maximally-localized_2018}.

 In this work we shall study in a holistic way both the effects of lattice deformation and the change in hopping due the change in alignment, which should be contrasted to related work in Refs.~\cite{AngeliMVACTF18,ChoiChoi18}. Our calculation of deformation bears some similarity to the work by van Wijk \emph{et al} \cite{van_wijk_moire_2014,wijk_relaxation_2015}, but the authors of those references mainly study a single layer on either bulk graphite or hBN. There are a few other papers that take a related approach \cite{dai_twisted_2016,uchida_atomic_2014,jain_structure_2017,huder2018electronic,yan2013strain,gargiulo2017structural} to lattice deformation, often in a slightly different context.
We start out by selecting a few modern potentials for graphene, and will analyze in detail the deformation of the bilayer systems. This will be validated by comparison to experimental results for strain solitons in bilayers, and will also be compared  to the results of a simplified  technique originally developed by Nam and Koshino \cite{nam_lattice_2017}. [We shall show in the Appendix that we can get a simple analytical series expansion for this model with minor modifications.] We then analyze the tight-binding model based on these data, and show that the lowest energy bands remain flat in the presence of a lattice deformation. Then we analyse a general way to extract a low-energy model from such data, and discuss potential issues there. In this work we concentrate on the study of lattice relaxation and electronic structure for a twisted sample at a fixed twist angle, $\theta \approx 1.05^\circ$. For this angle, the electronic properties depend on the choice of parameters used. In this respect, our analysis is rather different from those which select a given parametrization and modify the angle in order to obtain the narrowest band \cite{KV18,AngeliMVACTF18,Letal19}.   
Note, finally, that experiments determine the twist angles mostly from measuring the electron density required to fill the bands in the Moir\'e superlattice. The angles observed in this way need not coincide with the theoretically defined ``magic angles'' where the Fermi velocity at the $K$ and $K'$ points in the superlattice Brillouin Zone vanishes\cite{bistritzer_moire_2011}. Also,  it may make sense to use other definitions of the magic angles, such as those which lead to the narrowest bands, or to the largest gaps between the lowest states and the next ones.

Our goal is to quantify the uncertainty that exists in the basic description that is used as a starting point in most calculations of novel features in bilayer graphene. We shall not directly draw conclusions which approach is best; this should ideally be resolved by further experimental measurements of the Moir\'e structure and the local density of states, as should be accessible to STM measurements. We will, however, compare the lattice relaxation results to the measurements from Ref.~\cite{alden_strain_2013}, and show that we can obtain results that are rather comparable to the interface solitons seen in free-standing bilayers. 

Finally, we discuss the  robustness of results that rely on a particular Wannier function to describe superconductivity \cite{PZVS18,koshino_maximally-localized_2018,KV18,XB18,GZFS18,BV18,Detal18,YF18,PCB18,YV18,HYL18,XLL18,TCSS18}. We shall argue that the electron-assisted hopping model of Ref.~\cite{GuineaWalet18} looks like the most robust  way to obtain superconductivity, independent of the unknown details of the model. 

\section{Classical atomistic simulations}
In this section we shall investigate the deformation of free-standing graphene bilayers using atomistic potential models. We will employ a small number of well-established potential models, and for calculational simplicity we restrict our attention to those implemented in the LAMMPS package \cite{plimpton_fast_1995}.

For a single layer graphene, we shall use AIREBO-M \cite{oconnor_airebo-m:_2015} form of the AIREBO potential \cite{stuart_reactive_2000}, as well as the LCBOP-I potential \cite{los_improved_2005}, all of which work well for graphene. The reason we shall not use the AIREBO is its small equilibrium C-C spacing of  $1.397\,\text{\AA}$, unlike the standard value of $1.420\,\text{\AA}$ recovered for the AIREBO-M and LCBOP potentials. 
The nature of the interlayer interaction is a subtle question; the long-range and many-body nature of these potentials is discussed in Refs.~\cite{los_improved_2005, kolmogorov_registry-dependent_2005,van_wijk_moire_2014,leven_interlayer_2016, maaravi_interlayer_2017}. Most potential models are modifications of models first used for the interaction of graphene and HBN, and there is some indication that that this leads to a small underestimate of the corrugation of the graphene layers \cite{van_wijk_moire_2014}. In this work we shall only use the Kolmogorov-Crespi (KC) potential \cite{kolmogorov_registry-dependent_2005} and the interlayer potential (ILP) \cite{leven_interlayer_2016, maaravi_interlayer_2017}. 
Note that in the ``overlay" implementations of the ILP and KC potentials used in LAMMPS, the long range part of the AIREBO is switched off, effectively turning these potentials in re-parameterized REBO potentials \cite{brenner_second-generation_2002}. 

A different interlayer potential makes a difference in the results reported below. 
A detailed comparison between a large variety of choices in  Table \ref{tab:potential} of Ref.~\cite{rowe_machine_2018}, who derive a rather different form for the potential. 
In some of those more emphasis is placed on the vertical corrugation of bilayers (which is indeed important for the magnitude of interaction, and even though included in our work, may be slightly underestimated due to the nature of the potentials used). Others concentrate on strained graphene bilayers. The work by Jain \emph{et al} \cite{jain_structure_2017} employs a potential that is specifically designed for the out of layer deformation, but may be less well suited to the details of the in-layer deformation.  Nevertheless, this reference also contains an interesting discussion of the lattice deformation. Even though in Ref.~\cite{koshino_maximally-localized_2018} the importance of the corrugation is strongly emphasized, we shall argue that the in plane deformation of the lattice actually dominates when we take into account the subtleties of interlayer hopping in $AB$ stacking--rather than the pure two-body  form used in that reference.  Also, we expect vertical corrugation to be suppressed when the two layers are encapsulated within BN, as is the case in most
experiments. In all cases we expect the formation of $AB$ and $BA$ aligned regions separated
by domain walls (``interface solitons"). This problem is also discussed in Ref.~\cite{espanol_discrete_continuum_2018} using an analytic description of domain wall formation, but for rectangular domains.

We have performed simulations for a variety of supercell sizes, but will concentrate here on the case of a superlattice with periodicity $32\vec{a}_1+31\vec{a}_2$, with an angle between the two graphene lattices at the ``canonical value" of $1.05^\circ$, where we can also compare directly to the semi-analytical work by Nam and Koshino \cite{nam_lattice_2017}. This last approach is discussed in detail, in a simplified version that is susceptible to analytic solution, in the appendix.

We relax the lattice using a single supercell, with the dimensions chosen to contain
a graphene bi-layer lattice without deformation. We then relax, using a conjugate-gradient minimization, first the positions within flat layers, followed by a full relaxation of the carbon atoms. We have checked that these results do not depend on the method or specific order of relaxation used.

A useful way to analyze the in-plane deformation of the relaxed layers is to expand the new positions in terms of a lattice harmonics, 
\begin{eqnarray}
\vec{r}_{\parallel \vec i \sigma}&=& \vec{r}^{(0)}_{\parallel \vec i \sigma}
+ \vec u^\sigma(\vec{r}^{(0)}_{\parallel \vec i \sigma}),\\
\vec u^\sigma(\vec{r})&=&\sigma \vec u(\vec r)=\sum_{\vec q \in H_r} \sum_{j=0}^5 \frac{1}{i} \vec u_{R_{2\pi j/3}\vec q} e^{i (R_{2\pi j/3}\vec q) \cdot \vec r}.
\end{eqnarray}
Here $R_\theta$ is a 2D rotation over an angle $\theta$,  $H_r$ is the first sextant of the reciprocal lattice, i.e., the yellow domain in Fig.~\ref{fig:kpoints},  and $\sigma=\pm$ denotes the top (bottom) graphene layer. The vector $\vec r^{(0)}$ denotes the 
undeformed graphene position, and the parallel symbol means we only look at the in-plane component. Due to three-fold symmetry we find we only need to specify a fraction of the coefficients,
\begin{equation}
\vec u_{R_{2\pi j/3}\vec q}=R_{2\pi j/3}\vec u_{\vec q}.
\end{equation}
%In each case we find that the AIREBO potential does not give enough stability to a graphene layer; it tends to exhibit short-wavelength noise for the vertical displacement, even 
%at the equilibrium spacing. The fit to a lattice-harmonic expansion is also less than perfect, making the results in this case somewhat suspect. This is probably linked to the approach in \cite{wijk_relaxation_2015}, who adapt the KC potential to fit the graphene lattice spacing of the REBO potential. Since it is unclear how to do this reliably, we have not done so.

\begin{figure}
\includegraphics[width=8.5cm]{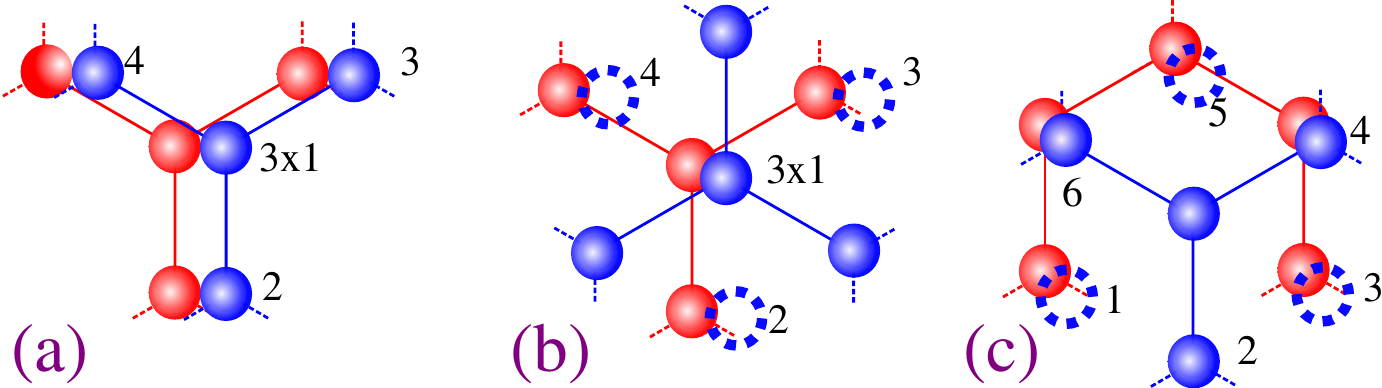}
\caption{Graphical representation of the terms used in Eqs.~(\ref{eq:wAA},\ref{eq:wAB}). The first diagram (a) is for  $AA$ alignment, the last two (b/c) define two situations in $AB$ alignment. The blue dotted circles are the positions of the blue upper layer carbon atoms inverted relative to the central one.}\label{fig:alignment}
\end{figure}

In order to compare the size of the $AA$ and $AB$ aligned domains, we construct a measure of alignment, by combining measures for $AA$ and $AB$ alignment. 
We first define the measure of $AA$ alignment by the function 
\begin{widetext}
\begin{equation}
w_{AA}(\vec r_{li})=\frac{1}{a^2}\delta_{\langle r_{li} r_{{\bar l}j}\rangle}
\left[3\left(\vec{r}_{li}-\vec{r}_{\bar lj}\right)^2+
\sum_k\left(\vec{r}_{lik+}-\vec{r}_{\bar ljk+}\right)^2\right].\label{eq:wAA}
\end{equation}
Here $l$ labels the layer, $\bar l$ denotes the opposite layer, $\langle r_{li} r_{{\bar l}j}\rangle$ denotes the atom $j$ closest to atom $i$ but in the other layer, and $\vec{r}_{\bar{l}jk\sigma}$ denote positions displaced over a single lattice spacing from $\vec{r}_{li}$ in the same layer, $\vec{r}_{lik\sigma}=\vec{r}_{li}+\sigma \delta^{(k)}_{li}$, where $\delta^{(k)}_{li}$, $k=1,2,3$ denotes the three nearest neighbors  $l$ of atom $i$.
In a similar way we define the quality of $AB$ alignment as 
\begin{eqnarray}
w_{AB}(\vec r_{li})&=&
\frac{1}{a^2} 
\min\left(\delta_{\langle l_i {\bar l}_j\rangle}
3\left(\vec{r}_{li}-\vec{r}_{\bar lj}\right)^2+
\sum_k\left(\vec{r}_{lik+}-\vec{r}_{\bar ljk-}\right)^2,\right.%\nonumber\\&&
\left.\sum_{k\sigma}\left(\vec{r}_{lik\sigma}-\vec{r}_{\bar{l}j}\delta_{\langle \vec{r}_{lik\sigma}, {\vec r}_{{\bar l}j}\rangle}\right)^2\label{eq:wAB}.
\right)
\end{eqnarray}
\end{widetext}
See Fig.~\ref{fig:alignment} for a graphical representation of these terms.
The factors of 3 in front of the terms involving the central atoms ensure that we use six atoms in every expression; they also weigh the central atom more heavily, when they are aligned. The value of $a$  is the graphene nearest-neighbor spacing.

We then use 
\begin{equation}
w=w_{AA}-w_{AB}
\end{equation}as a measure of alignment.  
We shall combine data from both layers in a single plot. Note that $w$ is extremal for perfect alignment, negative for $AB$ and positive for $AA$ alignment.
See Fig.~\ref{fig:alignment} for a graphical explanation of each of the terms.

\begin{figure}
\includegraphics[width=8cm]{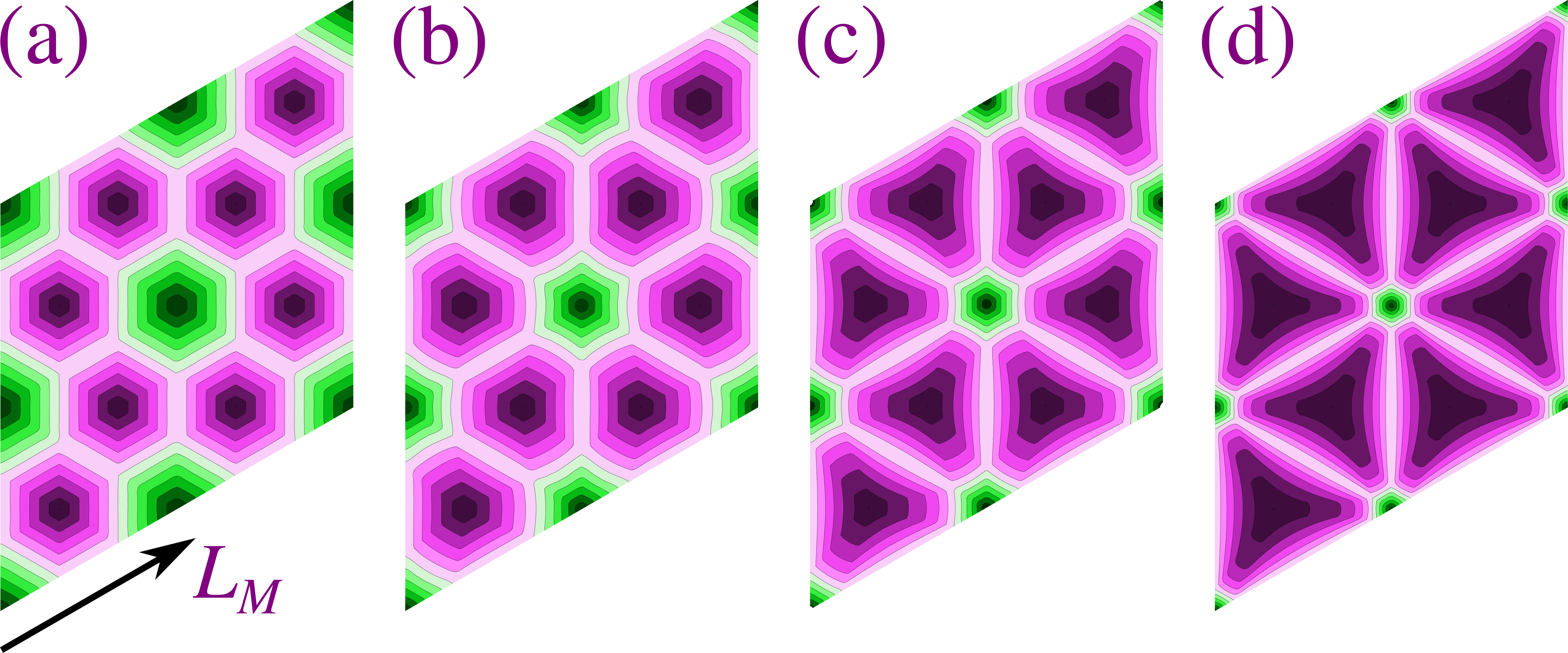}
\caption{Lattice  coordination for a lattice with sides $32\vec{a}_1+31\vec{a}_2$ lattice: a) no deformation, b) Nam and Koshino parameters, c) LCBOP+KC, d) AIREBO-M+ILP. In each case the colour map ranges from dark green for $AA$ alignment to purple for the $AB$ case. White indicates equal $AA$ and $AB$ alignment, and the scale is the same in all figures. Each of these figures shows \emph{four} unit cells, with AA registration at the corners and midpoints. }\label{fig:align}
\end{figure}

\begingroup
\begin{table*}
\caption{Lattice harmonic expansion of the deformation: the values of $\vec{u}_{\vec{q}}$ (in units of \AA \,) for the points labeled as in Ref.~\cite{nam_lattice_2017}.
%, rotated to agree with their definition of unit cell. \footnote{There seems to be a $\pi/3$ rotation mistake in the results of N\&K, which has been corrected in this data. 
%We have also divided there results by 2, since our form of $\vec{u}_{\vec q}$ is the translation of a point, rather than twice its value.} A dash indicates this value is too small to be  included in the calculation.
}\label{tab:potential}
\begin{ruledtabular}
\begin{tabular}{llllllll}
{$n_1,n_2$}  &{LCBOP+KC} &{AIREBO-M+KC} &{AIREBO-M+ILP} &{N\&K}\\
 (1,0) & (0.00042,0.04972) & (0.00141,0.07689) & (0.00129,0.07302) & (0.,0.02660) \\
 (2,0) & (0.00006,0.00323) &  (0.00025,0.01307) & (0.00026,0.01078) & (0.,0.00270) \\
 (2,1) &  (-0.0019,0.00347) &  (-0.00442,0.00809) & (-0.0051,0.00928) & (-0.00100,0.0017) \\
 (3,0) & (0.00001,0.00015) &  (0.00008,0.00272) & (0.00007,0.00191) & (0.,0.00036) \\
 (3,1) &  (0.00001,0.00001) &  (-0.00051,0.00182) & (-0.00075,0.00269) & (-0.00002,0.00035) \\
 (3,2) &  (-0.00005,0.00005) & (-0.00132,0.00153) & (-0.00189,0.00216) & (-0.00028,0.00020) \\
 (4,0) & (0.00001,-0.00002) &  (0.00003,0.00064) & (0.00002,0.00039) & -- \\
 (4,1) & (0.00004,-0.00014) &  (0.,0.00016) & (-0.00013,0.00071) & -- \\
 (4,2) &  (0.00003,-0.00006) &  (-0.0002,0.00038) & (-0.00042,0.00078) & -- \\
 (4,3) &  (0.00009,-0.0001) &  (-0.00015,0.00015) & (-0.00055,0.00053) & --
\end{tabular}
\end{ruledtabular}
\end{table*}
\endgroup

Clearly most of the results with a sensible in-layer potential (AIREBO-M and LCBOP-I) fall into groups that largely only depend on the interlayer potential: there are small differences, but they are much smaller than the effect of the interlayer potential. Also, the deformation of the Nam and Koshino analytic result is surprisingly small compared to what we find with modern potentials, with a pattern that appears to be somewhat different as well in the structure of the lattice harmonics, see Table \ref{tab:potential}. 
The best way to gauge the quality of these results is to look at the width of the strain solitons between the $AB$ and $BA$ regions. According to Ref.~\cite{alden_strain_2013}, this should be in the order of  $6\,\text{nm}$ for a shear boundary, which we believe  applies here. 
%For the $(32,31)$ case, we find that, as shown in Fig.~\ref{fig:soliton}, these results are essentially independent of the potential model: the peak around $10\,\text{nm}$ has a small width at half maximum of about $2\,\text{nm}$, a very sensible value, albeit smaller than the value ($6$) expected. 

%\begin{figure}[htp]
%\includegraphics[width=8cm]{soliton}
%\caption{A diagonal slice (bottom left to top right in Fig.~\ref{fig:align}) for a) Nam and Koshino data; b) LCBOP+KC; c) AIREBO-M+KC d) AIREBO-M+ILP. In each case the peaks around 12 and 25 nm are the inteface  solitons  between the BA and AB regions, with a width largely independent of the potential model.}
%\label{fig:soliton}
%\end{figure}
\begin{figure}
\includegraphics[width=8cm]{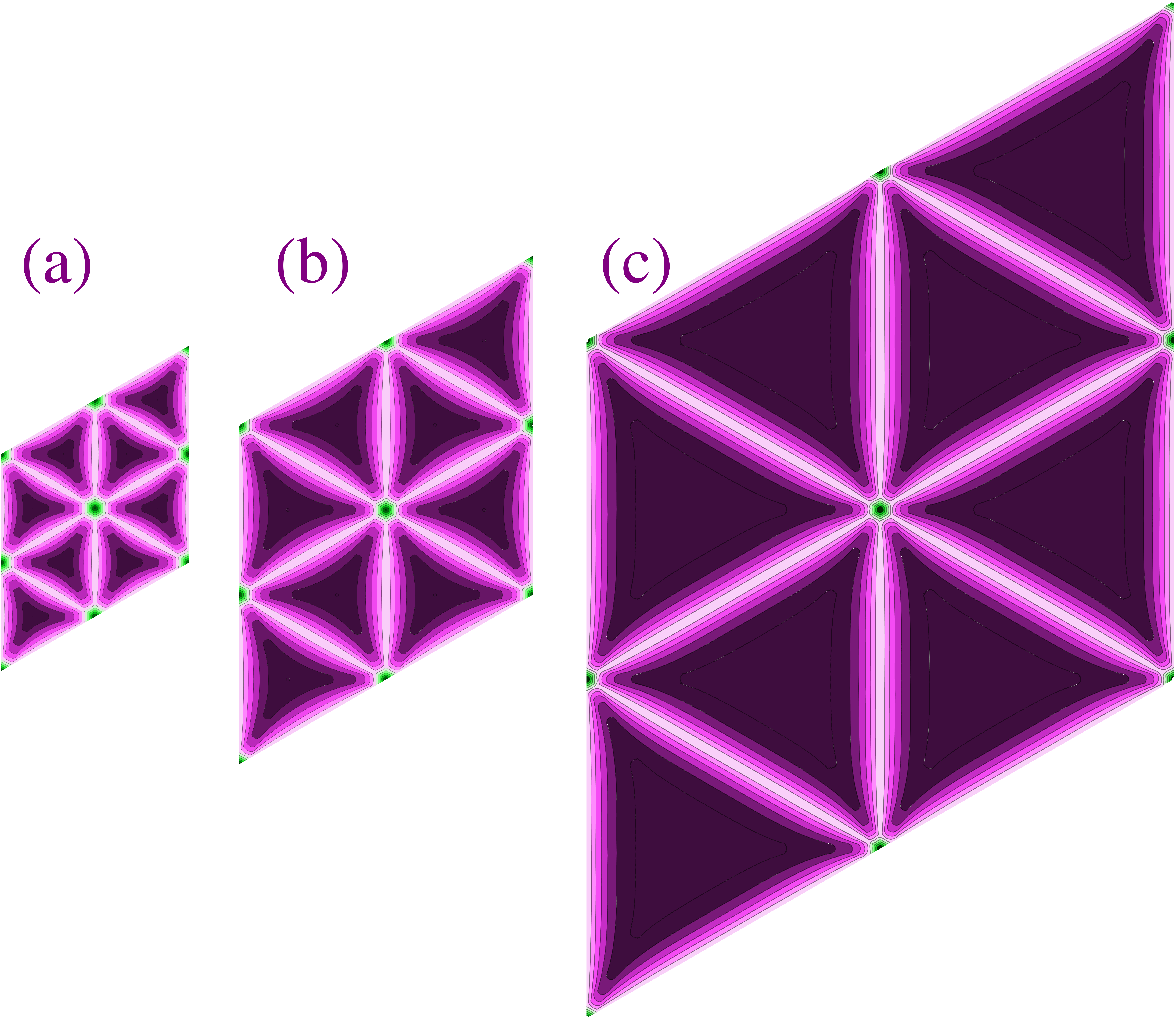}
\caption{Alignment for a $32\vec{a}_1+31\vec{a_2}$ (a), $50\vec{a}_1+49\vec{a_2}$ (b) and $100\vec{a}_1+99\vec{a_2}$ (c) bilayer graphene lattice described by the AIREBO-M+ILP potential.
}\label{fig:align2}
\end{figure}

\begin{figure}
\includegraphics[width=8cm]{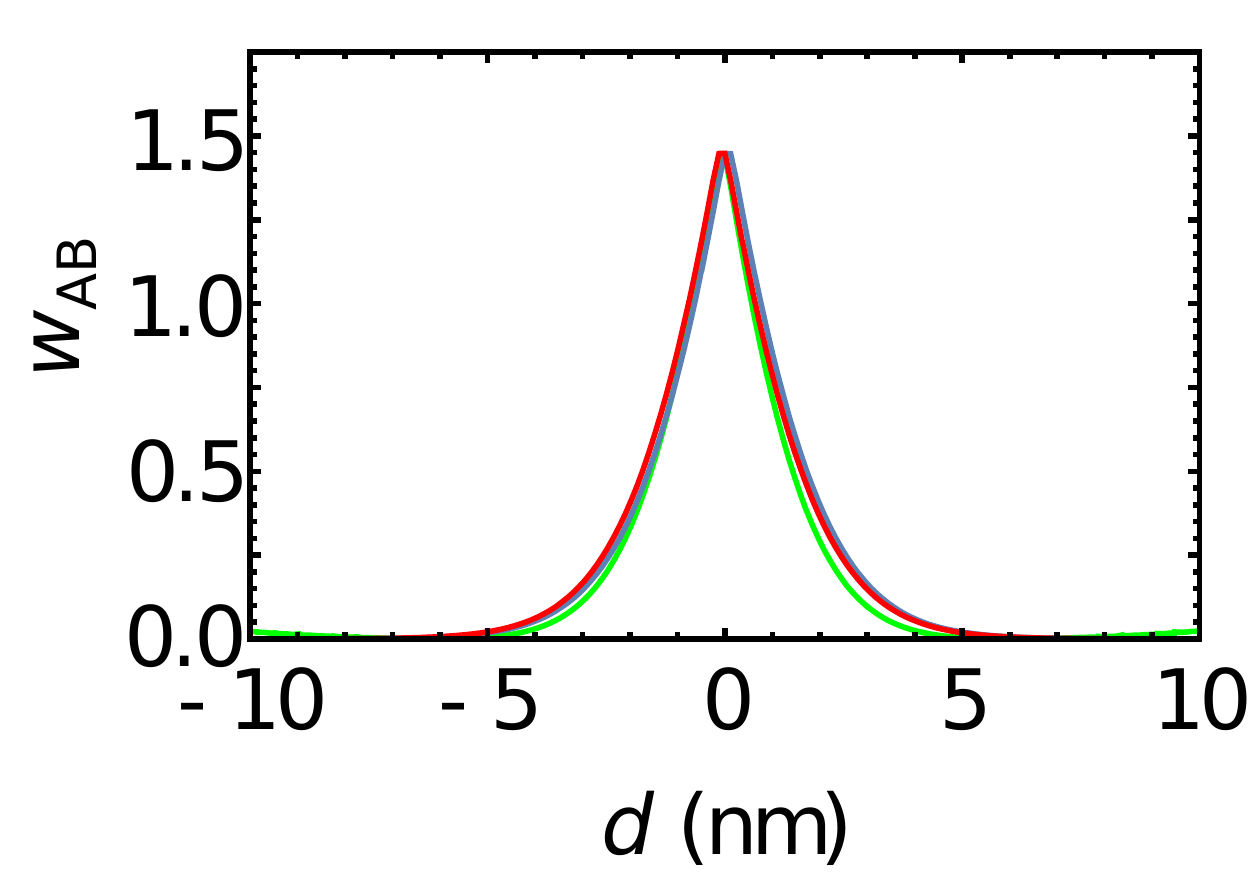}
\caption{Results for the interface soliton for a bilayer graphene lattice  with periodicity $50\vec{a}_1+49\vec{a_2}$ (green), $100\vec{a}_1+99\vec{a_2}$  (blue) and $150\vec{a}_1+149\vec{a_2}$  (red) described by the AIREBO-M+ILP potential.
In each case the strain soliton has a full width at half maximum of $2.3\,\text{nm}$.}\label{fig:soliton2}
\end{figure}

\begin{figure}
\includegraphics[width=7cm]{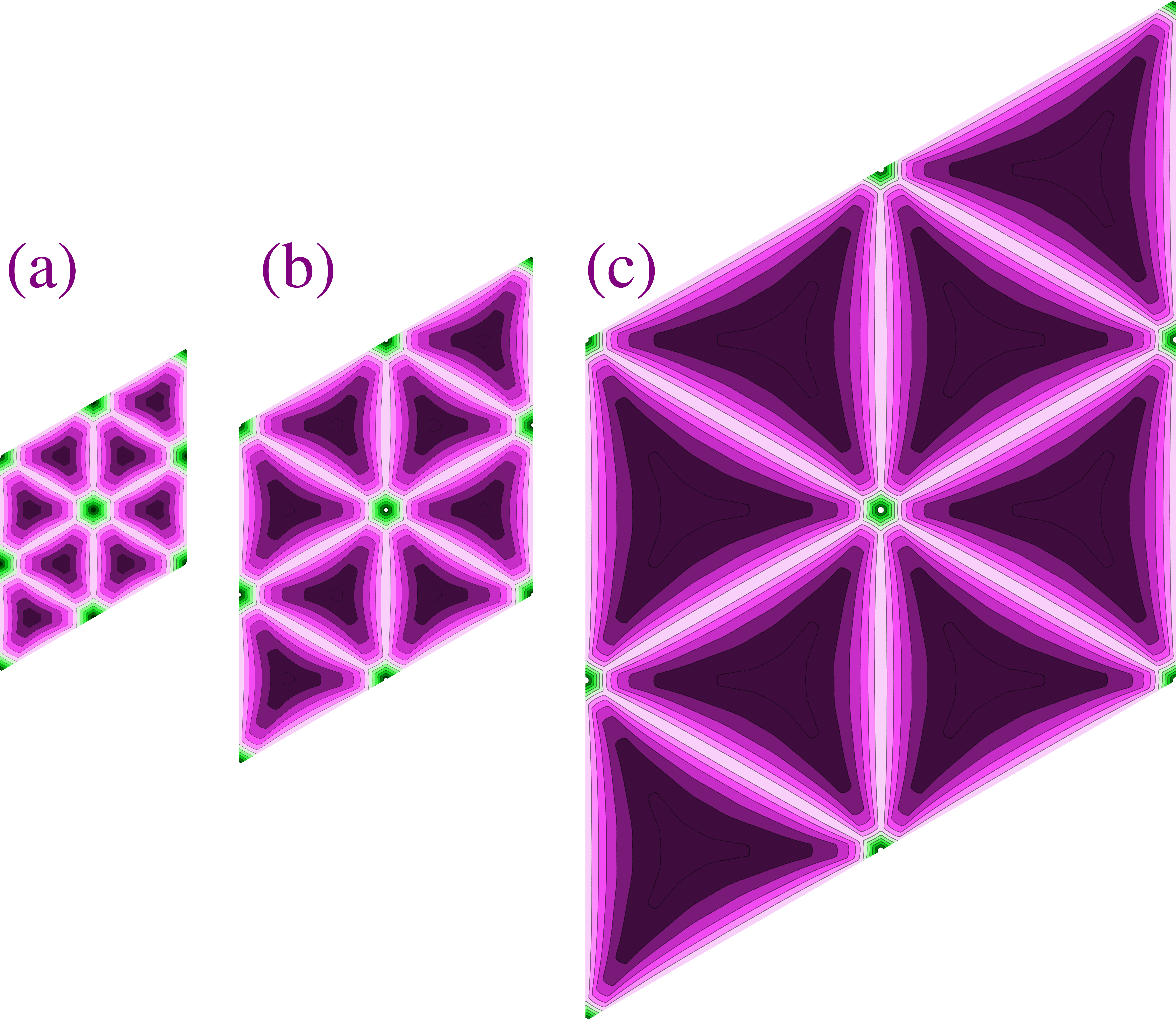}
\caption{Alignment for a $32\vec{a}_1+31\vec{a_2}$ (a), $50\vec{a}_1+49\vec{a_2}$ (b) and $100\vec{a}_1+99\vec{a_2}$ (c) bilayer  graphene lattice described by the LCBOP+KC potential.
}\label{fig:align3}
\end{figure}

\begin{figure}
	\includegraphics[width=8cm]{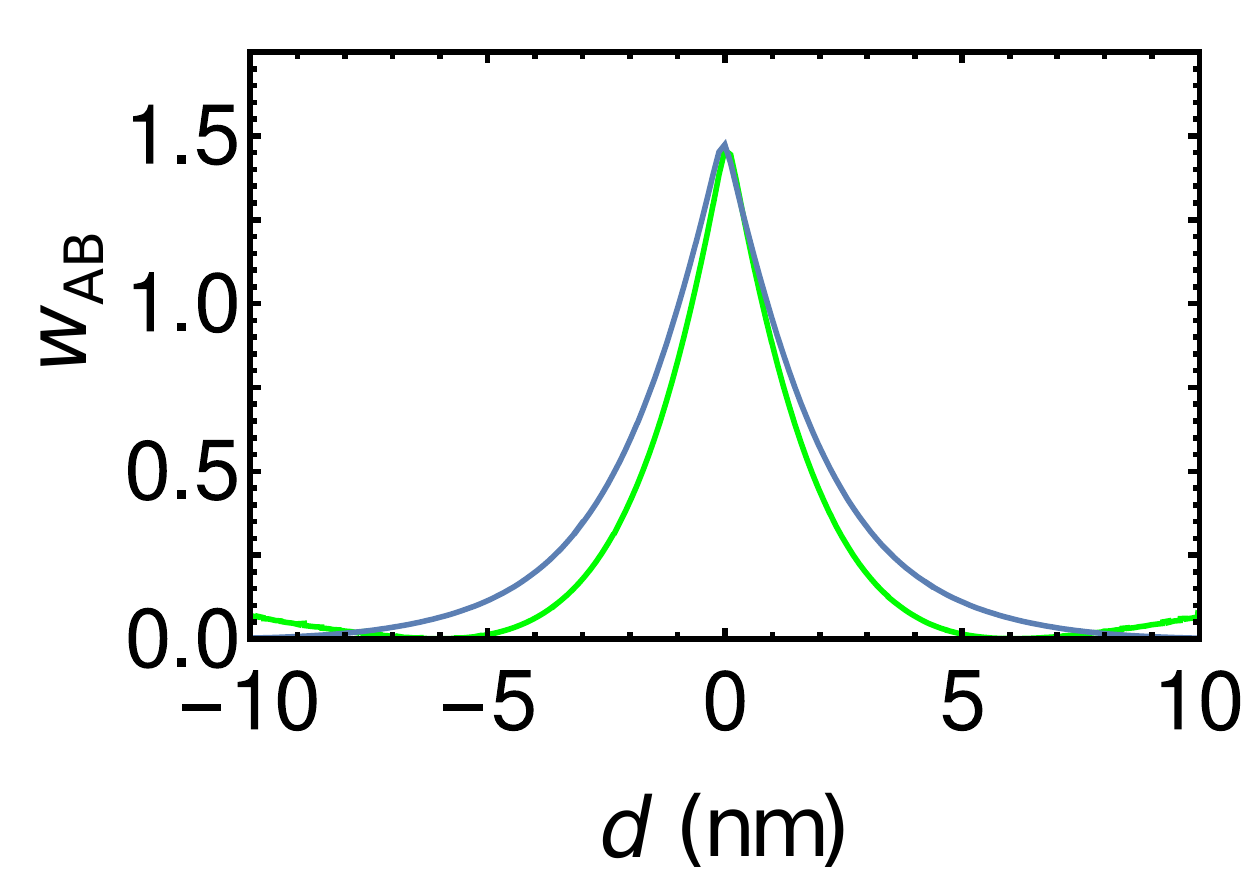}
	\caption{Results for the interface soliton for a $50\vec{a}_1+49\vec{a_2}$ (green), $100\vec{a}_1+99\vec{a_2}$  (blue) graphene lattice described by the LCBOP+KC potential.
		In the largest case the  soliton has a full width at half maximum of $3.1\,\text{nm}$.}\label{fig:soliton3}
\end{figure}

\begin{figure}
\includegraphics[width=8cm]{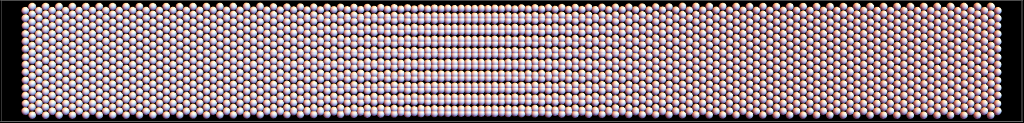}
\caption{The atomic positions in the soliton region (range as in Fig.~\ref{fig:soliton3}). This should be compared to Fig.~3A/B from Ref.~\citep{alden_strain_2013}.}\label{fig:L}
\end{figure}

In order to see whether we can reproduce such results, we need to look at larger domains (the ones studied in Ref.~\cite{alden_strain_2013} vary in size, but typical sizes seem to be at least of the order of $50\,\text{nm}$). As we can see in Figs.~\ref{fig:soliton2} and \ref{fig:soliton3} the size  of the interface soliton saturates, and we obtain values of the width that are in reasonable agreement with Ref.~\cite{alden_strain_2013}; a bit narrow for the AIREBO-M+ILC calculations, but rather similar to experiment for the LCBOP+KC ones. The latter case  also shows some slow growth with cell size, suggesting the results agrees even better with large-size results from Ref.~\cite{alden_strain_2013}.

As shown in Fig.~\ref{fig:L}, the pattern of atomic positions for the  the strain soliton looks very similar to that presented in Ref.~\citep{alden_strain_2013}. Of course our analysis is based on only on atomic positions, unlike the results in the paper cited, which are obtained either experimentally using an indirect measure of position, or described with substantial modelling of the probe from the position data. Nevertheless, the similarities are striking.

It is well-known from various simulations cited earlier that vertical corrugation of the graphene layers is important. We would expect a slight underestimate of the corrugation for our current choice of potentials. From potential models fitted specifically to reproduce deformation data \cite{jain_structure_2017}  we would expect a corrugation of about $d_{AA} = 0.360\,\text{nm}$  and
$d_{AB} = 0.335\,\text{nm}$. The values we find are $d_{AB}=0.335\,\text{nm}$ and $d_{AA}=0.351\,\text{nm}$ for the LCBOP+KC calculations, and $d_{AB}=0.336\,\text{nm}$ and $d_{AA}=0.356\,\text{nm}$ for the AIREBO-M+ILC one. This may show a small underestimate of the vertical corrugation, especially for the LCBOP+KC potentials. Since the $AA$ regions are very small, there is little sensitivity of the binding energy to the $AA$ distance, and thus small changes in the binding can have large effects on this distance without changing the in-lattice deformation and the energy balance appreciably.

\section{tight binding}
Having determined the atomic positions, we need to turn our attention to the electronic degrees of freedom, which we describe using a tight-binding approach.
We assume that the lattice unit of the Moir\'e superlattice is much larger than the graphene unit cell. Without deformation, the Moir\'e unit cell can then be divided into regions with $AA, AB$, and $BA$ stacking, which each occupy a similar fraction of the unit cell. tight-binding calculations suggest that the wavefunctions which describe the approximately flat bands near the neutrality point are then localized within the $AA$ regions\cite{TMM10}. 
Since this relies on many approximations, this deserves a detailed investigation.

We start from a tight-binding model for a single layer graphene  given by
\begin{equation}
H^{(l)}=t\sum_{\langle ij\rangle} c^{(l)\dagger}_i c^{(l)}_j+t'\sum_{\langle\langle ij\rangle\rangle} c^{(l)\dagger}_{i} c^{(l)}_j\,.
\end{equation}
Since we have allowed for deformation, we in principle have $t\rightarrow t_{ij}$. Since we shall concentrate on the intralayer coupling, and the changes in $t_{ij}$ are actually very small, we take $t_{ij}=t=\gamma_0=-2.7\,\text{eV}$\footnote{Please note that for some reason the value used in Ref.~\cite{koshino_maximally-localized_2018} is about $10\%$ smaller.}
 and for simplicity we shall use $t'=0$ (we have checked this makes no appreciable difference to our results). The fact that we use the same value of $t$ independent of lattice deformation is important: it means that the in-plane  wave functions, which only depend on the in-plane hopping parameters, are the same as those of the undeformed lattice; this simplifies the calculations, and is not a real restriction since bond-stretching is extremely small, as explained above.

We use three sets of interlayer hopping parameters; first of all a Koster-Slater exponential parametrisation 
\begin{equation}
t(r)=0.4 \exp(-a(r-r_0)),
\end{equation}
where we use $r_0$ as the flat-layer average distance as defined in Eq.~\ref{eq:r0}, which means that we cannot use this parametrisation for  the deformed lattices, since the $AB$ couplings become too strong due to the shorter inter-layer distance in the $AB$ regions. In principle, we could replace $r_0$ by the $AB$ distance, but since it is not clear that  this makes sense, we will not do so, but only apply this parametrisation for flat layers at an interlayer distance $r_0$.
(Note, however, that Ref.~\cite{koshino_maximally-localized_2018} appears to have carried out this program). 

\begin{figure}
\centerline{\includegraphics[width=\columnwidth]{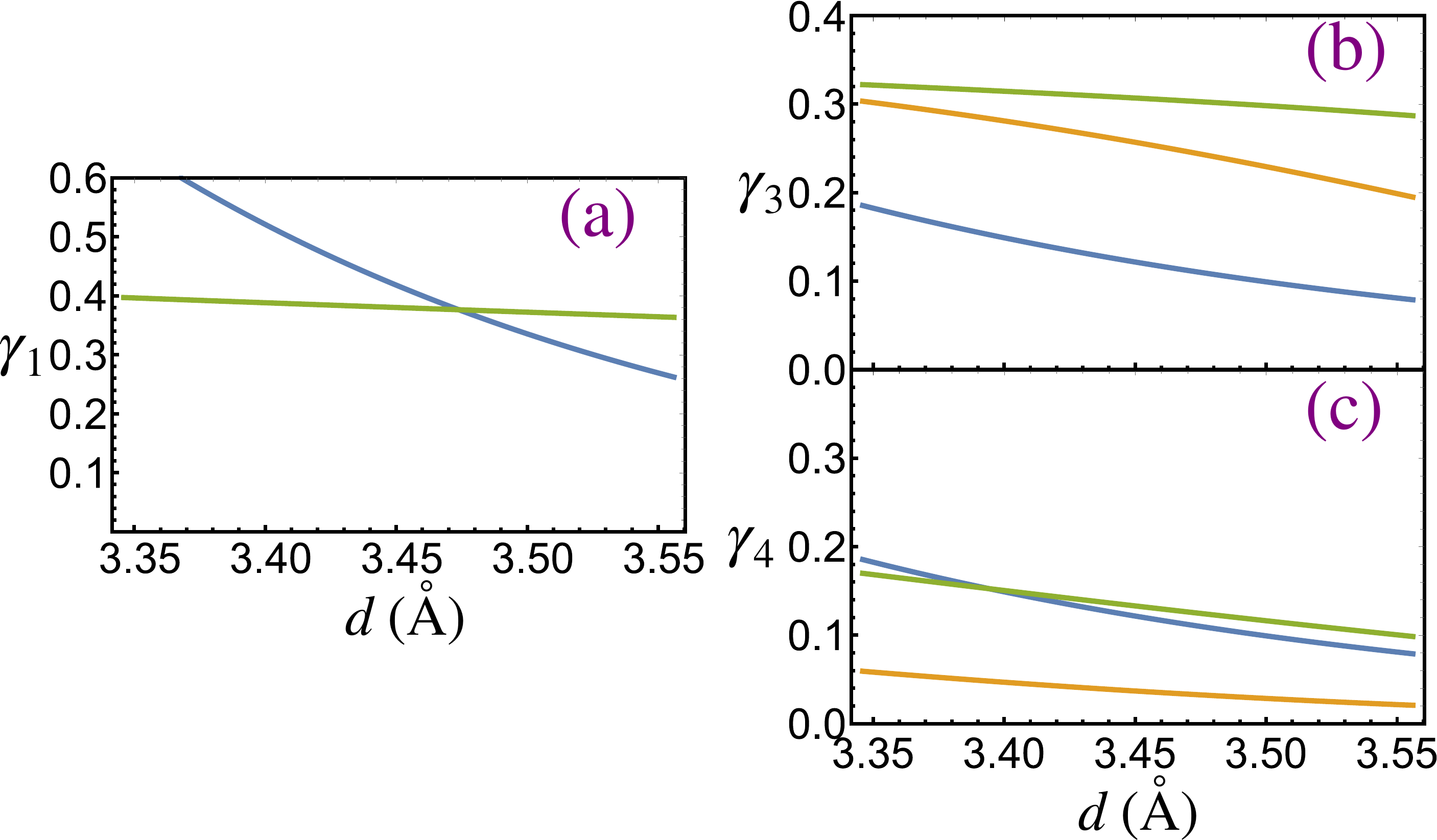}}
\caption{The values of the SWM $\gamma$ parameters for bilayer graphene in eV as a function of interlayer distance for each of our hopping parameters. The  blue line is the Koster-Slater parametrisation; the yellow line is the screened-1 hopping \cite{sboychakov_electronic_2015}, the solid green line is our screened-2 modification, see Table \ref{tab:pars}.  For the two-body Koster-Slater choice we always have $\gamma_3=\gamma_4$.}\label{fig:gammas}
\end{figure}

We use two sets of environmentally dependent (many-body) hopping parameters, both based on the work in \cite{sboychakov_electronic_2015}, who have designed a many-body screening function that is completely saturated by nearest neighbors only.
The form  we use is (with $\vec r=\vec r^{(2)}_2-\vec r^{(1)}_1$)
\begin{align}
V_1(\vec r^{(1)}_1,\vec r^{(2)}_2)&=\label{eq:scr1}
V_0 \left(\frac{|z_1-z_2|}{r}\right)^{\alpha_1}  
 \exp \left(-(\alpha_2 r)^{\alpha_3} \right)\nonumber \\&\quad\times(1-\tanh (\xi ))\,,\\
\xi&=\sum_{\vec{r}_3,l}f\left(\frac{|\vec r^{(l)}_3-\vec r^{(2)}_2|+|\vec r^{(l)}_3-\vec r^{(1)}_1|}{r}\right)\,,\\
f(x) &= \beta_1 \exp(-\beta_2 x^{\beta_3})\,.\label{eq:scr3}
\end{align}
We choose two sets of parameters; one, called ``screened-1", is essentially the parameter set from Ref.~\cite{sboychakov_electronic_2015} (with minor modifications); in the other one, ``screened-2",  a few parameters have been modified to even more closely represents the parameters in the bilayer SWM parametrisation as reported in Ref.~\cite{mccann_electronic_2013}.
The parameters for these two potentials are given in Table \ref{tab:pars}, and we study the behavior of the resulting $\gamma_i$ as a function of distance in Fig.~\ref{fig:gammas}. We see that our ``screened-2" potential only has a weak dependence on interlayer spacing, and gives $\gamma_1\approx 0.4 \text{eV}$, $\gamma_3\approx 0.3 \text{eV}$ and $\gamma_4\approx 0.1-0.2 \text{eV}$, in agreement with the values quoted in \cite{mccann_electronic_2013}. The original screened potential is essentially identical for $\gamma_1$, has a slightly smaller and more variable $\gamma_3$, and has
a value of $\gamma_4$ more appropriate to graphite. The Koster-Slater coupling has a great sensitivity to the interlayer spacing, which is especially problematic for $\gamma_1$, and follows the relation $\gamma_3=\gamma_4$, where $\gamma_3$ is rather small. Also, it has a 6-fold symmetry for the couplings near $\gamma_3$, whereas only a threefold symmetry is present.

\begin{table}
\caption{Table of parameters used in Eqs.~(\ref{eq:scr1}--\ref{eq:scr3}) for our two many-body screened hopping models.}\label{tab:pars}
\begin{ruledtabular}
\begin{tabular}{lll}
parameters & screened-1\cite{sboychakov_electronic_2015} & screened-2\\
\colrule
$V_0$  &1.06191 eV &1.06191 eV\\
$\alpha_1$ &0.476 & 1.0\\
$\alpha_2$ &0.295 \AA$^{-1}$& 0.295 \AA$^{-1}$\\
$\alpha_3$ &1.411 & 1.411 \\
$\beta_1$ &6.811 &6.811 \\
$\beta_2$ &0.01 & 0.01\\
$\beta_3$ &19.176 & 20.5\\
\end{tabular}
\end{ruledtabular}
\end{table}

We use two sets of deformation parameters, LCBOP+Kolmogorov-Crispi (LKC) and AIREBO-M+ILP (AILP). We also study the effect of fixing the separation, keeping the in plane deformation. For such a flat layer, as might be more appropriate when graphene bilayers are each mounted on HBN, we fix the separation of the layers at an average value of \begin{equation}
r_0=3.460\,\text{\AA}\,.\label{eq:r0}
\end{equation}
For the case studied here (with a unit vector of $32 \vec{a}_1+31\vec{a}_2$, and a Moir\'e angle of $1.05^\circ$) the distance in the deformed lattices is given in Table \ref{tab:dist}.
\begin{table}
\caption{Lattice displacements  for each of our classical potential models}.\label{tab:dist}
\begin{ruledtabular}
\begin{tabular}{llll}
model &max ($AA$) & min ($AB$) & mean\\
\colrule
LKC&$3.506$\,\AA&$3.347$\,\AA&$3.378$\,\AA \\
AILP&$3.556$\,\AA&$3.378$\,\AA&$3.398$\,\AA
\end{tabular}
\end{ruledtabular}
\end{table}

\begin{figure*}[p]
\begin{center}
\includegraphics[width=0.8\textwidth]{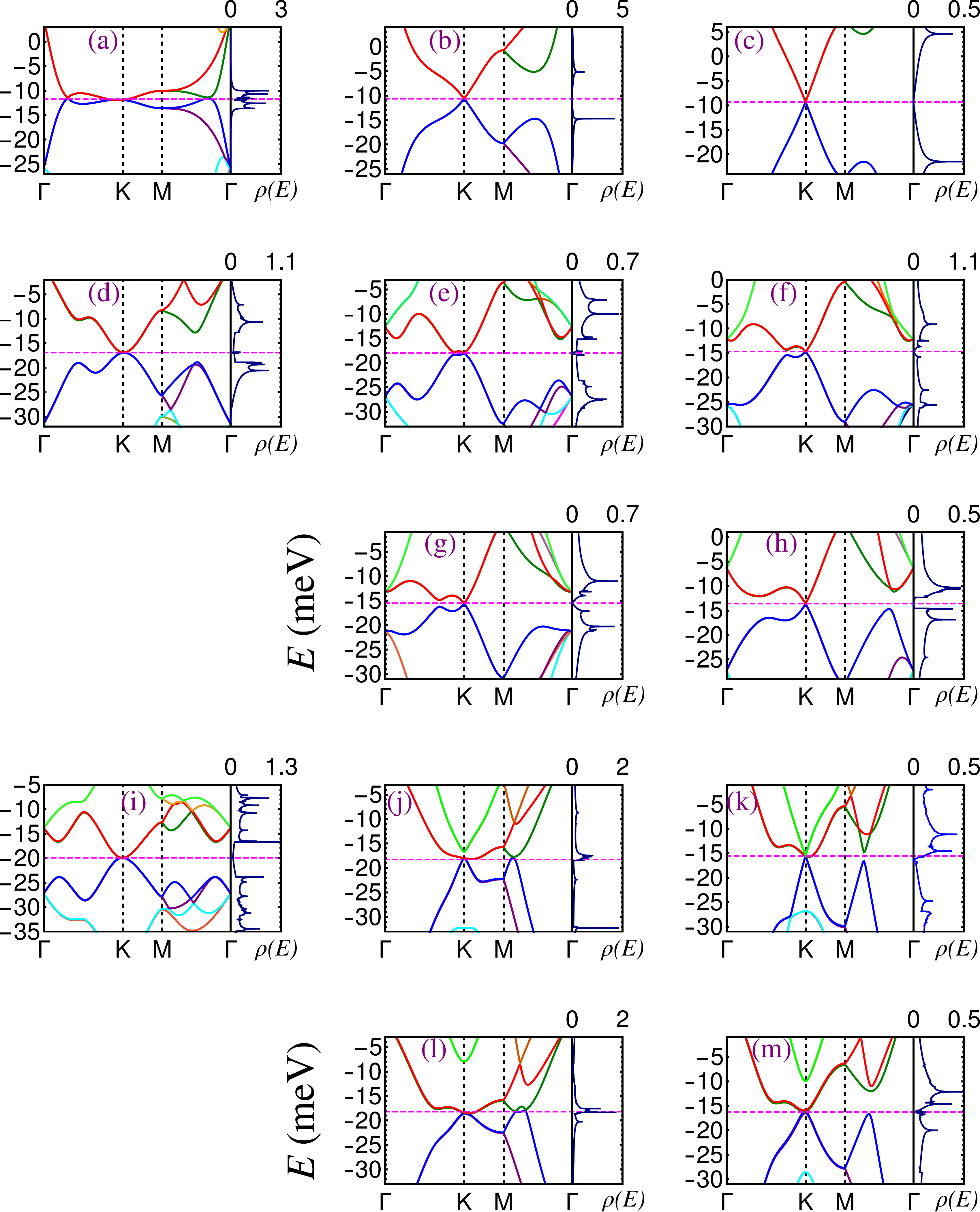}
\end{center}
\caption{Band structure (left) and density of states (right) of a Moir\'e commensurate superlattice of lattice parameter $\vec{ b}_1 = 32 \vec{ a}_1 + 31  \vec{ a}_2$. The twist angle is $\theta \approx 1.05^\circ$. All of these figures have a constant nearest-neighbor in-layer coupling.
(a)  Undeformed lattice with an exponential Koster-Slater  inter-layer coupling;
(b) LKC deformed lattice with the Koster-Slater coupling;
(c) AILP deformed lattice with the Koster-Slater coupling;
(d) Undeformed lattice with our screened-1   inter-layer coupling;
(e) LKC deformed lattice with the  screened-1 coupling;
(f) AILP deformed lattice with the screened-1 coupling;
(g) LKC deformed lattice without vertical corrugation with the  screened-1 coupling;
(h) AILP deformed lattice  without vertical corrugation with the screened-1 coupling;
(i) Undeformed lattice with our screened-2   inter-layer coupling;
(j) LKC deformed lattice with the  screened-2 coupling;
(k) AILP deformed lattice with the screened-2 coupling;
(l) LKC deformed lattice without vertical corrugation with the  screened-2 coupling;
(m) AILP deformed lattice  without vertical corrugation with the screened-2 coupling.}
\label{fig:bandsnd}
\end{figure*}

In Fig.~\ref{fig:bandsnd} we analyze the effect on the spectrum from both the deformation and interlayer coupling.
Again, we only show results for a superlattice
%with unit vector $\vec{b}_1 = 32  \vec{ a}_1 + 31  \vec{ a}_2$, where $\vec{ a}_1$ and $\vec{ a}_2$ are the unit vectors of one of the graphene lattices. The 
twist angle  $\theta = 1.05^\circ$,  where the length of the superlattice unit vector is $L_M=| \vec{ b}_1 | \approx 134.2\,\text{\AA}$, and the unit cell contains $11908$ carbon atoms. 
%We notice that for the extreme short-range interaction we have extremely flat states; 

%For the most realistic many-body screening terms we find a small but interesting band splitting, and if we increase the range of the interaction (by removing the many-body screening) we get an unrealistically large splitting. 

All of these results are for a regular bilayer without deformation. So what is the effect of deformation? 
%Since the interlayer potentials used in the previous section all contain many-body effects comparable to the screening used in the environment-dependent hopping parameters, we concentrate on the most realistic case (but only for the standard $(32,31)$ lattice).

As we can see in Fig.~\ref{fig:bandsnd}a, for an undeformed flat lattice and the Koster-Slater hopping parameters, we indeed get flat bands. There also is a secondary Dirac point, so we have probably gone a little bit beyond the first  magic angle, which for this interaction is slightly larger. Both of the environment-dependent potentials are a bit too long-range for flat layers, and leads to a larger spitting, but still of the order of $40\,\mathrm{meV}$. Adding lattice deformation leads to much more complicated spectra; secondary Dirac points appear in many places, and culminate in the complicated spectra seen in Fig.~\ref{fig:bandsnd}j-m. These still have a high density of states near the Fermi energy, mostly in a range of $\pm 5,\text{meV}$, so are likely to be susceptible to superconducting instabilities.
None of these show a  gap between the ``flat bands" and the remaining states at the $\Gamma$ point. This will be investigated further below, but it seems unavoidable with the strength of the SWM parameters required, unless we look at a larger twist angle: Whenever we have a second Dirac point for the flat-band calculation, we find that bands touch at the $\Gamma$ point. 

The tight-binding models shown here, even though for the canonical angle, show that this is not the magic angle as determined by the band structure. Since our results should at least be close to those by Koshino \emph{et al} \cite{koshino_maximally-localized_2018}, who find clear flat bands and a gap, we first investigate what effects reducing the interlayer coupling and reducing the in-layer Fermi velocity have (these authors use a 10\% reduced Fermi velocity). Note that experimental STM data seem more consistent with the larger bandwidth, and the larger Fermi velocity\cite{Ketal18,Cetal19} ,(see also related capacitance measurements in\cite{Tetal19}).

\begin{figure}
\includegraphics[width=\columnwidth]{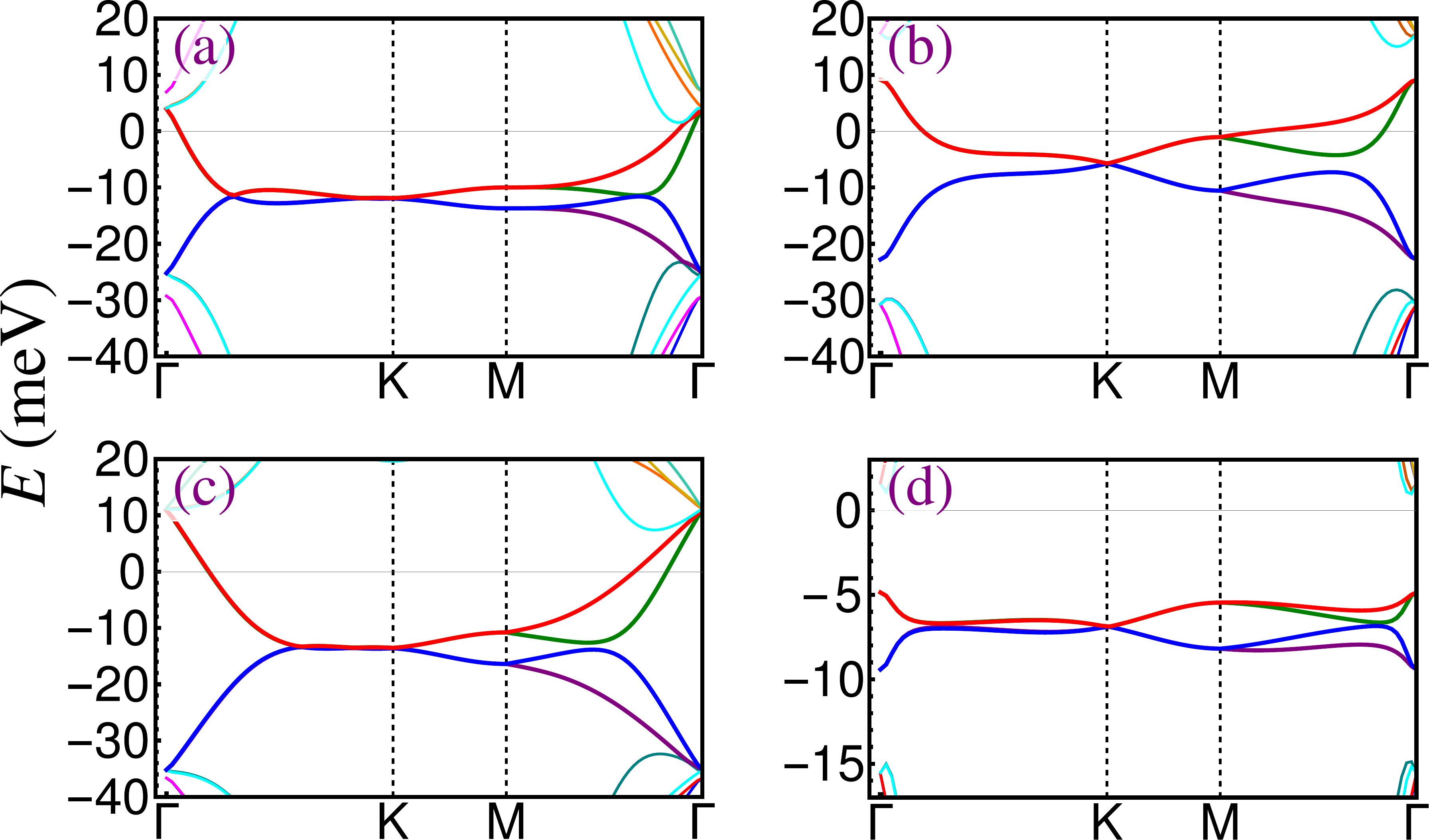}
\caption{Bands in an undeformed graphene bilayer for a Koster-Slater coupling. a) Is for our choice of hopping parameters; b) shows the effect of reducing the Fermi velocity by 10\%; c) shows the effect of reducing the interlayer hopping by 10\% and d) shows the effect of both changes simultaneously.\label{fig:exp_scaled}}
\end{figure}
As can be seen in Fig.~\ref{fig:exp_scaled}, we see that the most important effect is the scaling of the Fermi velocity, as introduced in Ref.~\cite{koshino_maximally-localized_2018}, which removes the secondary Dirac point, and opens a gap at the $\Gamma$ point. A reduction in only the interlayer hopping has almost the same effect, but the secondary Dirac point still remains. This also means that no gap opens at the $\Gamma$ point, where a degeneracy remains. Finally, making both changes has an effect that seems very similar to the bands studied elsewhere. 
%We leave it to the reader to estimate how realistic such a model really is; 
From the discussion in this paper, it should become clear that at an angle of $1.05^\circ$ this is not the behavior seen;  
note that only a model with a gap at $\Gamma$ gives the possibility to project on the 2-band Wannier states.

Clearly we could have reached a similar result by choosing a larger alignment angle; again, keep in mind that all calculations have been done at the same angle, but note that the combination of interlayer couplings and Fermi velocities means that in many cases our chosen twist angle is smaller than  the first magic angle for those parameters.

We believe that the in-plane deformation is crucial; the out of plane deformation is likely to be suppressed  by the encapsulation of the graphene layer by BN.
%Thus we would argue that any continuum model needs to reflect this behavior, and also that it may be less than clear-cut what a good continuum model looks like, even though we would prefer the results from the LCBOP+KC potentials. 

\section{Continuum projection}
Most of the work on studying bilayer graphene has been done using the continuum model, using a $\vec k\cdot \vec p$ model expanded around the a point halfway between the nearest layer, i.e.,  graphene, Dirac points \cite{bistritzer_moire_2011}.
 In most cases a simple symmetric model is used; the main exception is the work of Koshino \emph{et al} \cite{koshino_maximally-localized_2018}, where the effect of the rippling of the graphene layers is
used to modify the coupling strength in the $\vec k\cdot \vec p$ model, but with a simple two-body Koster-Slater interlayer hopping only. As explained in the previous section, we probably under-estimate the rippling, but our results also include the effects of lattice deformation and the many-body screening in the hopping, which have a much stronger effect.
\begin{figure}
\includegraphics[width=4cm]{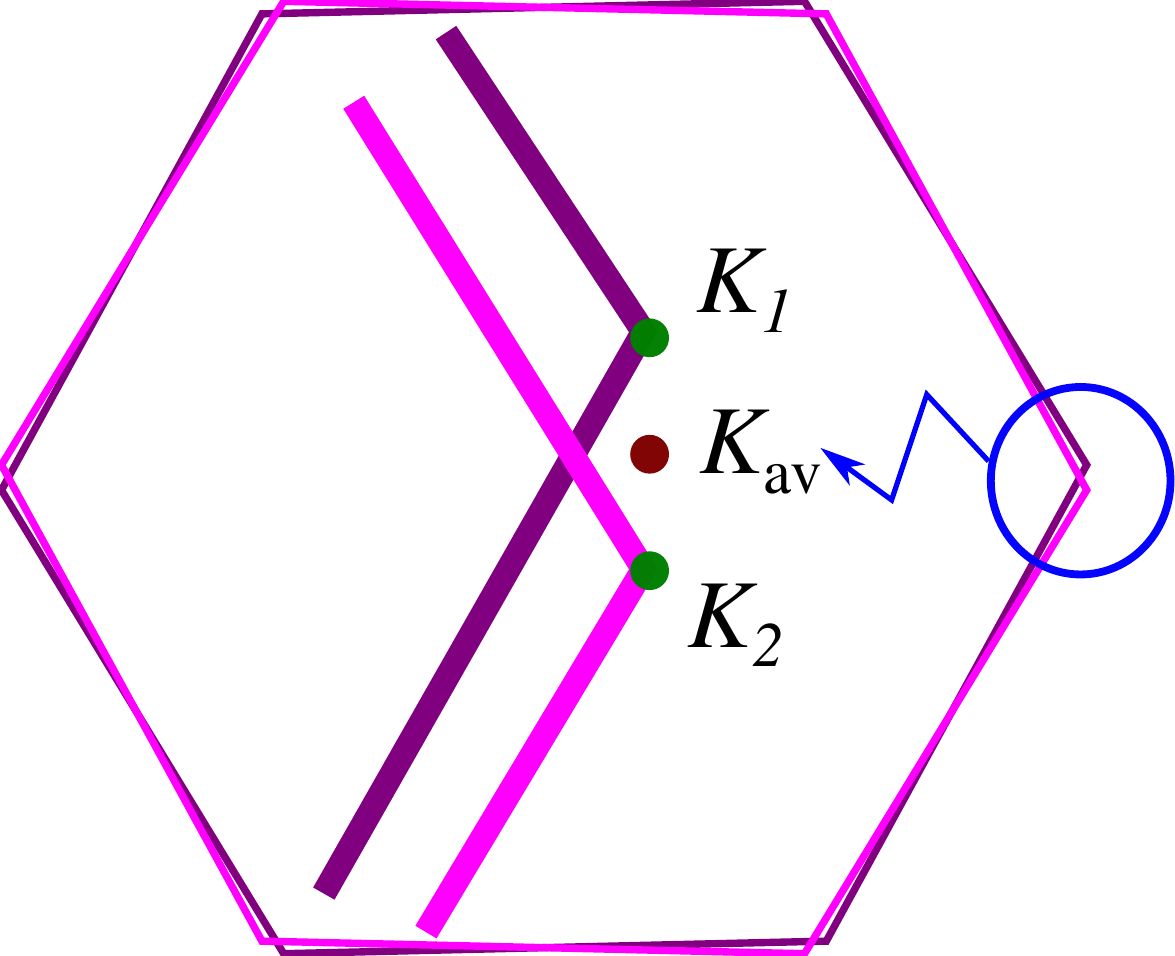}
\caption{The $K$ points of the two layers, with the expansion point in the middle.
These fold onto the $\bar{K}$, $\bar{K}'$ and $\bar{M}$ points.} \label{fig:expand}
\end{figure}

In order to avoid confusion, we shall denote the bilayer's first Brillouin zone points by a bar in the following; unbarred quantities refer to the single layer graphene points. 
The technique is straightforward, if a little confusing at first. We refer to Fig.~\ref{fig:expand} for a graphical representation of the edge of the Brillouin zones of the graphene lattice. Since these are slightly twisted, the reciprocal space is also not perfectly aligned, and the $K$ points in the two layers, $K_1$ and $K_2$, no longer coincide. For small angles these points are relatively close together, and develop a continuum Hamiltonian around the point $\vec{K}_\text{av}$ halfway between the two $K$ points. On folding to the bi-layer graphene Brillouin zone these $K$ points map to \emph{inequivalent} points $\bar{K}$ and $\bar{K}'$ in the bilayer-superlattice Brillouin zone. The point $\vec{K}_{\text{av}}$ maps to $\bar{M}$.
Since the Fermi velocity of graphene is rather large, we expect that only momenta near these two Dirac points play a role. To make that more precise, we expand the bilayer wave functions in products of the states of the graphene layers. The fact that the Moir\'e pattern is periodic means that only states that differ in momentum by the superlattice reciprocal lattice vectors mix. More precisely, we write for a single electron state of momentum $\vec k$ in the $p$th band:
\begin{align}
\ket{\bar{\vec{k}}p}&=\sum_{\vec n,s_1}c^{(p)1}_{\vec n,s_1}(\vec k)\ket{\vec k+\vec{G}_{\vec n}, s_1}_1\otimes\ket{0}_2
\nonumber\\&
+
\sum_{\vec m,s_2}c^{(p)2}_{\vec m,s_2}(\vec k)\ket{0}_1\otimes\ket{\vec k+\vec{G}_{\vec m}, s_2}_2.\label{eq:bilayer}
\end{align}
Here we choose for convenience $\vec{k}$ as the ``unfolded" momentum corresponding to the momentum $\bar{\vec k}$ in the FBZ, i.e., the equivalent momentum nearest the two Dirac points, and $s_l$ is a sublattice index for each layer. The states $\ket{\vec k,s_l}_l$ are the standard plane wave solutions (since we have not modified the in-plane hopping parameters, the positions used here are the undeformed lattice positions)
\begin{equation}
\braket{\vec r}{\vec k,s_l}_l=\frac{1}{\sqrt N}\sum_{\vec{r}_{sl}} e^{i\vec k \cdot \vec r}\delta(\vec r- \vec r_{sl}),
\end{equation}
where $\vec{r}_{sl}$ are the positions in sublattice $s$ in layer $l$.
The expression \eqref{eq:bilayer} is exact, and only becomes approximate on restriction
of the superlattice sums. Before we do that, we first look at at the representation of the tight-binding Hamiltonian in this basis.

Clearly we can write a block diagonal form
\begin{equation}
H=\begin{pmatrix}
 H_{11} & H_{12}\\
H_{21} & H_{22}
\end{pmatrix}\,,
\end{equation}
where each block itself is a block of $2\times2$ matrices in sublattice space, with the dimension determined by the number of vectors $\vec{G}_M$ included. Thus
 $(H_{11})_{\,\vec m s_1,\vec m' s'_1}=\epsilon_{\vec k+ \vec G_{\vec m},s_1}\delta_{\vec m\vec m'}\delta_{s'_1\bar{s}_1}$ and $(H_{22})_{\,\vec n s_2,\vec n' s'_2}=\epsilon_{\vec k+ \vec G_{\vec n},s_2}\delta_{\vec n\vec n'}\delta_{s'_2\bar{s}_2}$. For small $\vec n$ and $\vec m$ this is a slightly modified Dirac Hamiltonian (see below). The off-diagonal terms do allow coupling between different momenta due to the periodic Moir\'e, and the allowed couplings are of the form
\begin{align}
H_{12}&=\prescript{}{1}{\bra{\vec k +\vec{G}_{\vec m}}} H(\vec k) \ket{\vec k +\vec{G}_{\vec n}}_2\,.
\end{align}
The momentum dependence of the tight-binding Hamiltonian originates from the imposition of periodic boundary conditions, and also from the fact that, even though short-ranged, the interlayer potential, $V_{12} ( \vec{ r}_1 , \vec{ r}_2 )$, where $\vec{ r}_1$ and $\vec{ r}_2$ reside in different layers, is non local. 

The standard continuum approximation makes the assumption that the interlayer potential is only significantly different from zero if $| \vec{ r}_1 - \vec{ r}_2 | \ll L_M$, where $L_M$ is the Moir\'e lattice unit. Then, the position dependence of the interlayer potential should be well approximated by \begin{equation}V_{12} ( \vec{ r}_1 , \vec{ r}_2 ) \approx V_{12} \left( ( \vec{ r}_1 - \vec{ r}_2 ) / 2 \right)\,.\end{equation} When expressed in momentum space, this approximation neglects the dependence on the average momentum, $\vec{ k} + ( \vec{ G}_m + \vec{ G}_n ) / 2$. Even though in most cases considered here this is a small effect, the gaps we observe are also very small, and we would like to take a more careful approach

We will make the approximation
$H(\vec k)=H(\vec k+(\vec{G}_{\vec{m}}+\vec{G}_{\vec{n}})/2 )$. This is rigorously true only if $(\vec{G}_{\vec{m}}+\vec{G}_{\vec{n}})/2$ is a superlattice vector. Nevertheless, we write $\vec{ K} =\vec{ k} + ( \vec{ G}_m + \vec{ G}_n )/2$ and $\vec{ \kappa}= ( \vec{ G}_m - \vec{ G}_n )/2$ and
\begin{align}
H_{12}&=\prescript{}{1}{
\bra{ \vec{K}-\vec{\kappa}/2}
}
H(\vec{ K}) 
\ket{\vec{ K}+\vec{ \kappa}/2}_2\nonumber\\
&=U_{s's}(\vec{ K},\vec{ \kappa})
\,.\label{eq:Hovl}
\end{align}

The basic idea of the continuum model \cite{LPN07,bistritzer_moire_2011}, see also \cite{koshino_maximally-localized_2018}, is that for low energy states, and thus momenta near the Dirac points, we can make the approximation that the dependence on the average momentum can be replaced by the momentum at the point $\vec{ K}_\text{av}$ halfway between the two $K$ points.
This would  mean that for momenta near $\vec{K}_\text{av}$ we only consider the following quantity
\begin{equation}
U_{s's}(\vec{ K},\vec{ \kappa})\approx U_{s's}(\vec {\bf \kappa})
=U_{s's}(\vec{ K}_\text{av},\vec{ \kappa})
,\label{eq:Hovl2}
\end{equation}
which is slightly more satisfying approach to the local-potential approximation.
Since the interlayer coupling usually  falls of quickly with momentum,  $U$ is thus dominated by a few points on the triangular $\vec{G}_{\vec m}$ lattice \cite{bistritzer_moire_2011}. Actually, for reasons not perfectly clear to us, it seems better to use $\vec K_\text{av}=\vec K_1$ for a low-order truncation to $U$--this preserves the three-fold symmetry normally imposed on the model. We find that even that is not the optimal approximation, as is shown below.

\begin{figure*}
\includegraphics[width=0.9\textwidth]{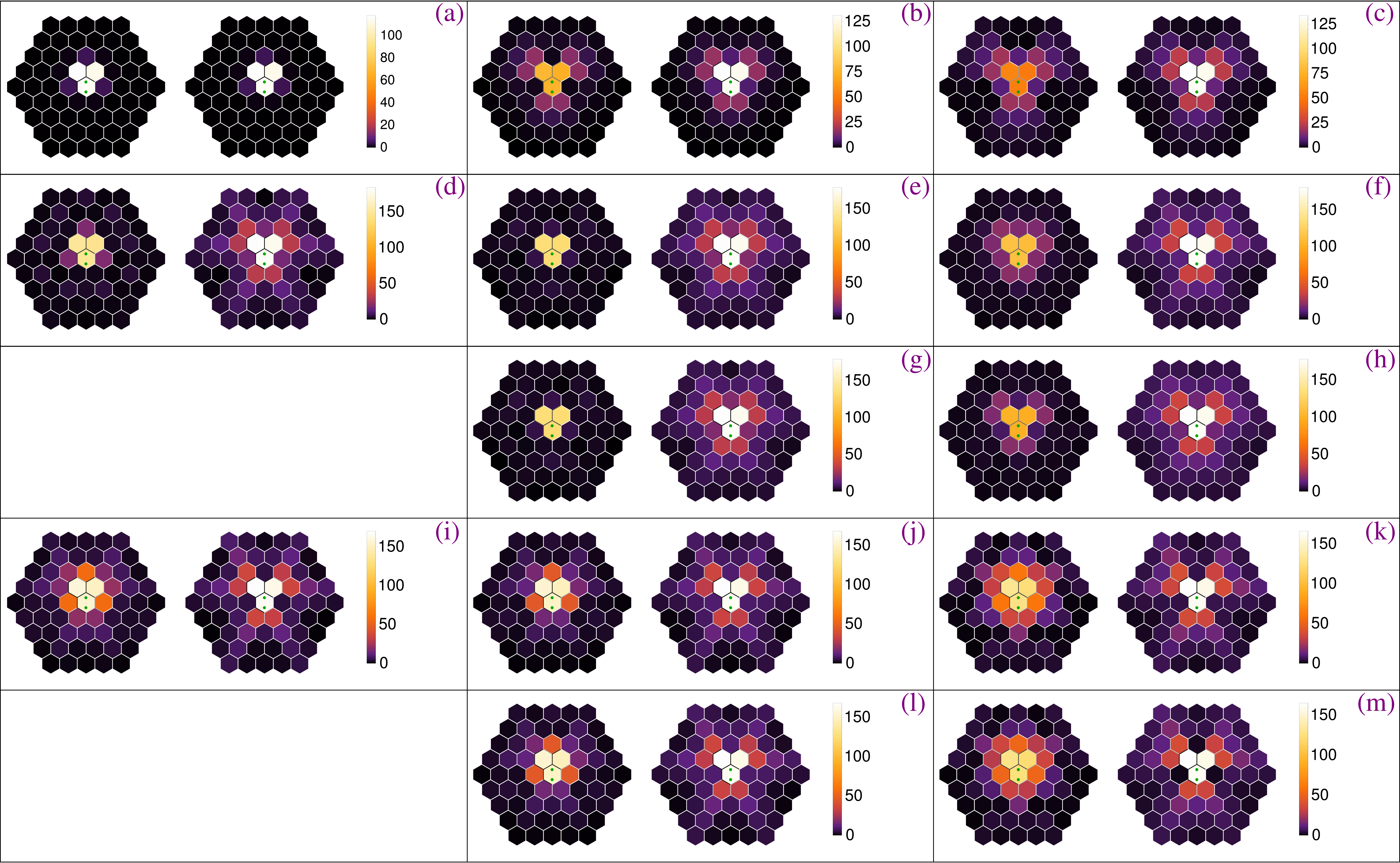}
\caption{The magnitude of the matrix elements \eqref{eq:Hovl} as a function of the momentum transfer $\vec{k}$. Each hexagon--or rather its midpoint--denotes a single superlattice vector $\vec{G}$, and the color shows the absolute value of the relevant matrix element. The green circles are the points $\vec K^{(1)}$ and $\vec K^{(2)}$. 
The plots  correspond to the spectra shown in Fig.~\ref{fig:bandsnd}, and are labeled accordingly.
 In each case the entries on the left are $AA$ couplings, and the ones on the right the $AB$ ones. Note that the color-scale used is non-linear to better show differences between small matrix elements.}\label{fig:overlap}
\end{figure*}

Let us first look at what these matrix elements \eqref{eq:Hovl} are for the problems studied previously; we study all of the cases shown in Fig.~\ref{fig:bandsnd} in Fig.~\ref{fig:overlap}. We indeed find that for a Koster-Slater potential and a flat lattice the couplings are dominated by 3 wave vectors (which is the model underlying Refs.~\cite{bistritzer_moire_2011,koshino_maximally-localized_2018}). We clearly see  that in all cases the three nearest-neighbor vectors dominate but that the decay is slower both due to lattice deformation and the change of the interlayer hopping parameters.  For the $AA$ coupling we always find a small asymmetry between the $\vec{G}=0$ coupling and the other two strong couplings (by a few percent), removing some of the symmetries of the model, which can be restored, see below.
For an undeformed graphene lattice and the Koster-Slater hopping parameters (a), the parameters are essentially those quoted in Ref.~ \cite{bistritzer_moire_2011}, after a small rescaling of the strength. The $\gamma_3$--$\gamma_4$ asymmetry in the remaining results clearly has  a big impact. For an undeformed lattice (b/c), we see larger $AB$ than $AA$ couplings, but there is an indication that the coupling decays slightly more slowly, and  some additional couplings may be thus be required in the continuum model. For the relaxed and deformed lattices, we find a more substantial difference between the $AA$ and $AB$ couplings, where the $AA$ coupling is smaller (by $15-30\%$) than the $AB$ one. It should come as no surprise that the AILP results, which have the smallest $AA$ regions, show the largest difference.

What we have not shown is the imaginary parts: normally one assumes that the coefficients in $U$ are real after removing a trivial phase-dependence. In our case they seem to develop small but significant imaginary parts.

We shall now apply the expansion of $U$ in two different approaches: Since, due to the large energy cost associated with moving up the Dirac cones, only momenta $\vec K$ near $\vec K_{\text{av}}$ will contribute, it is usually considered sufficient to replace the average momentum dependence  by the central value,  and expand the graphene dispersion to linear order about this same point.
The second idea is based on the fact that we can do better at little cost: for the momenta that are relevant, a linear approximation of dependence of $U$ on $\vec K$ (expanded near $\vec{K}_\text{av}$), can easily be combines with the full in-layer dispersion. 
We truncate the matrix diagonalization to the $n$th hexagon, and we find that a projection with $n=3-5$  (depending on the range of $U$) is sufficient to reproduce the energy of the flat bands, which is similar to the truncation proposed in the literature; we usually use a few more hexagons to ensure convergence.
Slightly more concerning is the effect of an expansion of the Dirac Hamiltonian about the $K^{(1,2)}$ points. In the most complete calculation we  use the exact dispersion, described by the off diagonal element of the in-plane Hamiltonian
\begin{equation}
\tilde{\bf t}_{ND} (\vec k)=t \left|2 e^{\frac{i a k_y}{2}} \cos \left(\frac{1}{2} \sqrt{3} a k_x\right)+e^{-i a k_y}\right|\,.
\end{equation}
%\begin{table}
%\caption{\label{tab:table1} Expansion at the $\Gamma$ point}
%\begin{ruledtabular}
%\begin{tabular}{ccccc}
%& full & expansion at $k$ & expansion at $K_\text{av}$\\
%\hline
%$E_1$ & $-24.5$& $-27.6$& $-110$\\ 
%$E_2$ & $-24.4$& $-21.7$& $-62$\\
%$E_3$ & $3.7$& $5.3$&$-2.1$\\
%$E_4$ & $3.6$& $2.7$&$-9.9$\\
%\end{tabular}
%\end{ruledtabular}
%\end{table}

\begin{figure*}
\includegraphics[width=10cm]{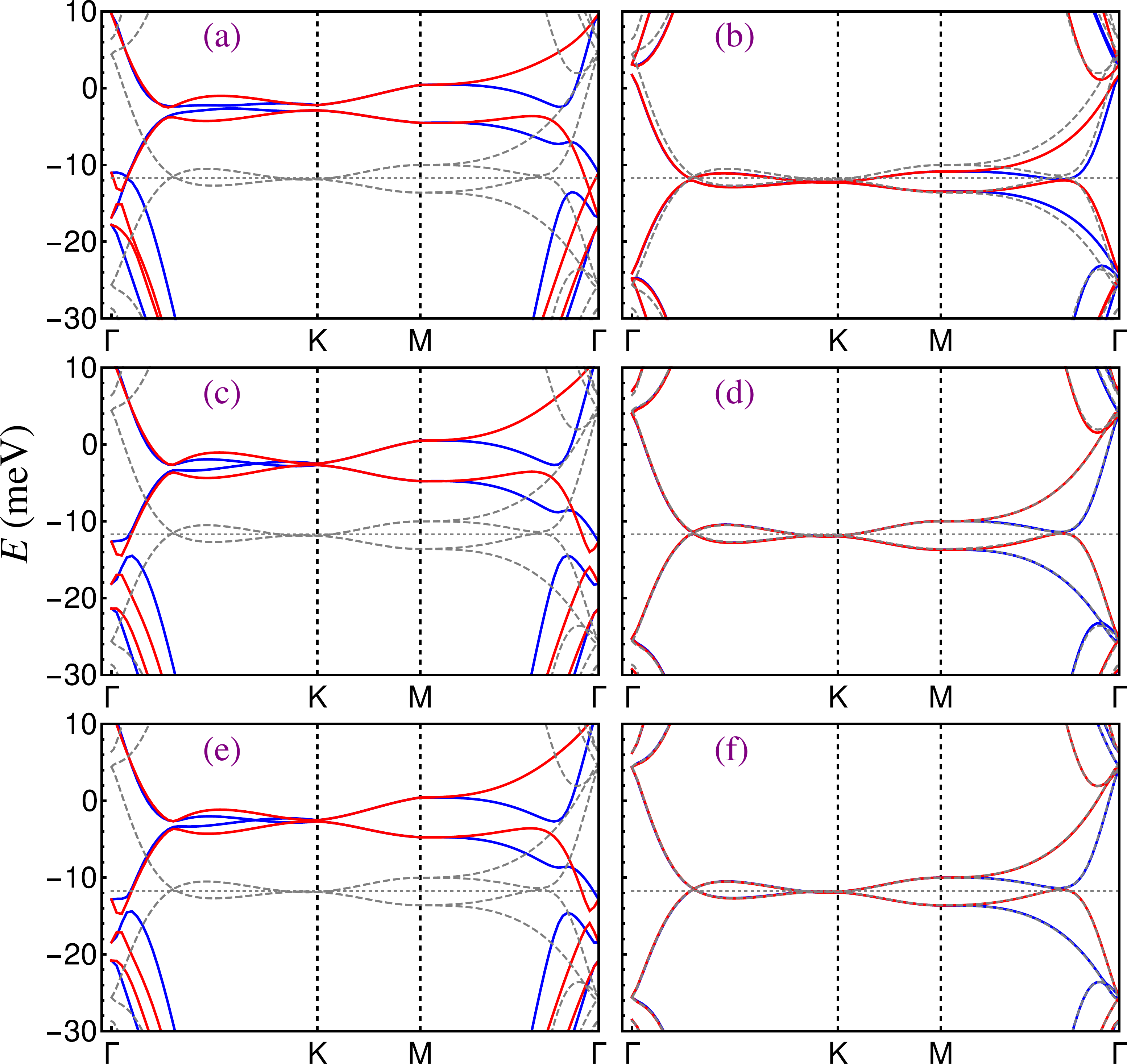}
\caption{Spectrum for various form of the continuum model for the case of the SWM model without lattice deformation. (a) is the ``standard" Bistritzer-MacDonald truncation, with only three interlayer matrix elements and Dirac in-layer dispersion; (b) is the same model now with the in-layer tight-binding dispersion, and the full $k$ dependence of the interlayer matrix elements. (c) and (d) are similar figures, but now including the next group of intralayer matrix elements as well; (e) and (f) finally includes all the matrix elements that give non-perturbative effects. The red and blue curves are the two valleys of the model; the gray lines are the exact diagonalization.}\label{fig:undefco}
\end{figure*}

In Figs.~\ref{fig:undefco} and \ref{fig:lkcco} we give two examples of calculations for two extreme cases; a complete set is shown in the supplementary material. 

Let us look at the ``standard case", Fig.~\ref{fig:undefco} first. we see a rapid convergence of the results with the range of $U$; the three dominant matrix elements are almost sufficient. We see a small symmetry breaking along the $\Gamma$--$M$ lines for the Bistritzer-McDonald calculation. This could have been avoided by replacing $K_\text{av}$ by $K_1$, and we would get the correct degenerate spectrum, but not the particle-hole asymmetry--i.e., the Fermi energy is incorrect. Interestingly enough, by using the linear $K$-dependence of $U$ and the full dispersion (and both are required) we get a perfect reproduction of the tight-binding spectrum.

The situation gets much more interesting for Fig.~\ref{fig:lkcco}, which corresponds to Fig.~\ref{fig:bandsnd}j. Clearly even for this complicated case the full calculation converges to something close to the tight-binding results (we could have added probably one more hexagon  of couplings, which would have converged). The standard approximation, based on just three harmonics, gives a rater poor approximation.

We now present some results for the continuum projection (a complete set can be found in the supplementary material). We selected two cases of most interest: the SWM model without deformation and the AILP+KC deformation, as probably the most reasonable cases to investigate.

As we can see in Fig.~\ref{fig:undefco}, the inclusion of a full dispersion and the the dependence of the intralayer matrix elements is required to get the degeneracy of the energies in the two valleys along the line $\Gamma$-$K$. 
\begin{figure*}
\includegraphics[width=10cm]{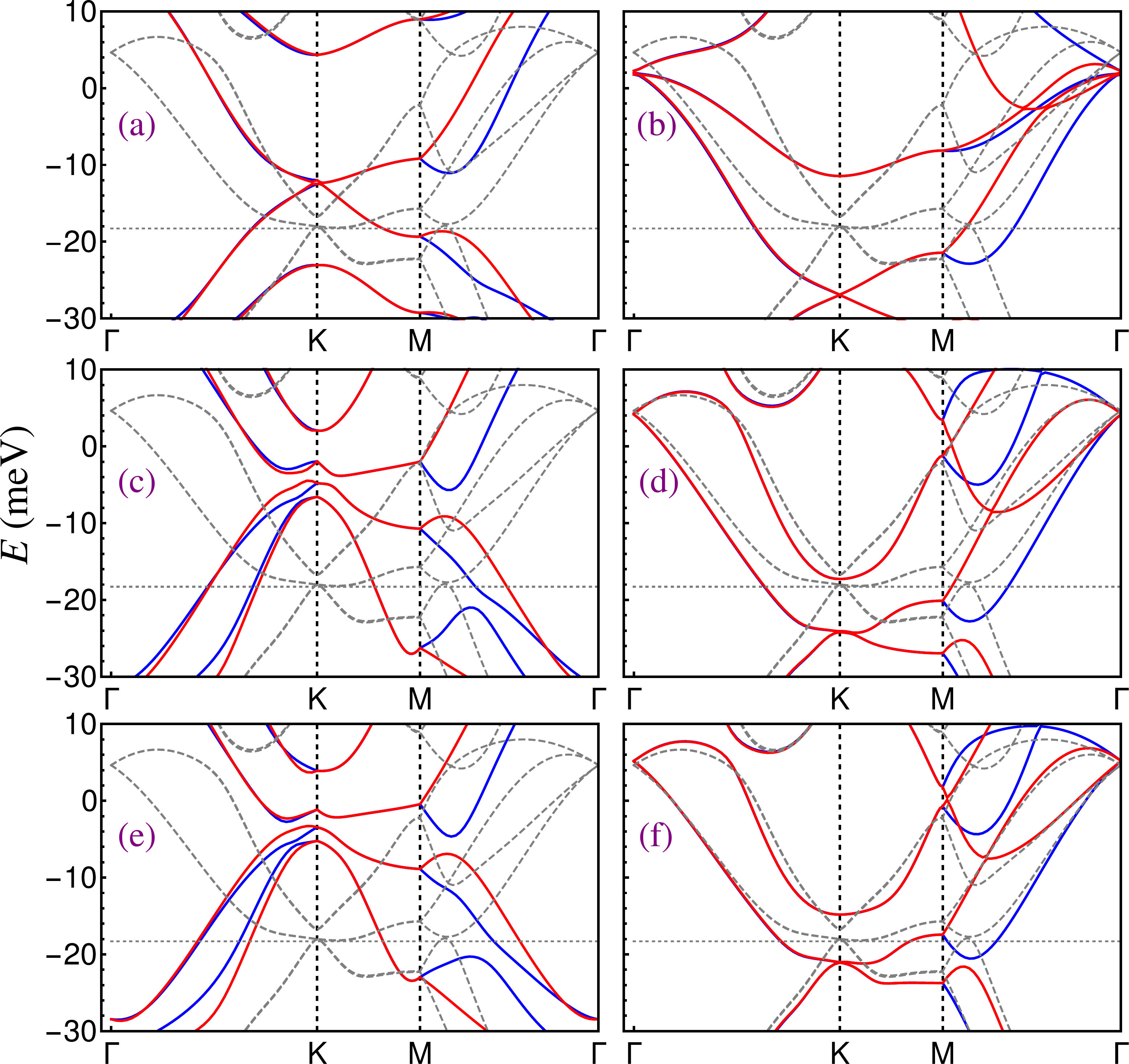}
\caption{Spectrum for various form of the continuum model for the case of the second SWM model with LKC lattice deformation. See Fig.~\ref{fig:undefco} for details of the results presented.}\label{fig:lkcco}
\end{figure*}

This changes in spectrum will clearly also have important consequences for the wave function--which in turn can be used to construct the Wannier functions. These are shown in the supplementary material.

\section{Conclusions.}
We have presented a comprehensive analysis of the lattice relaxation in twisted graphene bilayers, and its effect on the electronic properties, due to the modulation of the interlayer hopping. Calculations have been carried out for a Moir\'e superlattice with lattice vector $\vec{ L}_M = 32 \vec{ a}_1 + 31 \vec{ a}_2$, where $\vec{ a}_1$ and $\vec{ a}_2$ are the unit vectors of the graphene lattice. The twist angle is $\theta \approx 1.05^\circ$. Note that our approach is complementary to other studies, where one selects the angle which gives the narrowest bands near the neutrality point, and keeps the parametrization used fixed\cite{KV18,AngeliMVACTF18,Letal19}.

The relaxation is calculated using classical interatomic force models, and the electronic states are determined using tight-binding models. We have compared different force models, and different dependencies of the interlayer electronic hopping parameters on atomic positions, and find rather similar results.
The relaxed positions of the atoms are used as input for the calculation of the electronic structure, calculated using tight-binding models. Different parametrizations of the couplings are used: i) hoppings between orbitals in different layers which combine a form factor which reflects the symmetry of $p$ orbitals, and a simple exponential dependence on distance, and ii) hoppings that depend on the distance {\it and} the local environment of the two orbitals involved in the process. Models of type ii) reproduce the difference between the SWM parameters $\gamma_3$ and $\gamma_4$ needed to describe aligned bilayers and graphite. For a fixed twist angle, $\theta \approx 1.05^\circ$, the low energy bands show a significant dependence on both the range of the interaction and whether the hopping parameters depend solely on interatomic distances, or they also include other features of the environment. To some extent, the results can be interpreted as a parameter dependent shift of the ``magic angle'', where the low energy bands are narrowest. When the choice of parameters is such that the magic angle is greater than $1.05^\circ$, we find new band crossings and Dirac points\cite{HLSCB19}.

The low-energy electronic bands show a significant dependence on the amount of lattice relaxation and on the dependence of the interlayer hopping parameters on distance and local environment. The bandwidth of the lowest bands at neutrality is probably a bit larger than for the non-relaxed case, but still has a large density of states within a few meVs. However, a number of features, such as the number and location of additional band crossings (Dirac points) and saddle points (van Hove singularities) varies considerably as function of the model being used. The overlap, or lack thereof, of the lowest bands and neighboring bands is also quite sensitive the choice of parameters, within a range of physically sensible ones. 

Finally, we have studied the connection between tight-binding and continuum $\vec{k} \cdot \vec{p}$ models. We find that the number of harmonics required in a continuum approximation is dependent on the strength of the lattice relaxation and details of the interlayer hopping, but that effective continuum models can be defined in all cases.

We have analyzed the minimal continuum models required to approximate the electronic bands obtained from tight-binding calculations defined at the atomic scale. The complexity of the continuum models depends significantly on the range of the hoppings, and on whether they depend significantly on the local environment. Isotropic couplings which do not decay too abruptly with distance are reasonably described with the standard model based on an expansion with three harmonics of the interlayer hoppings. A continuum description is possible for all tight-binding models considered, although more than three harmonics are required in some cases, especially when the hopping parameters depend on the local environment. Even with a large number of harmonics, such a continuum model can be an effective way to study a tight-binding model, especially when also adding residual interactions. This is of course dependent on the method for extracting the coupling parameters from a tight binding model. This calculation can be done quite simply, and only relies on the construction of a tight-binding Hamiltonian, not its diagonalization.

We have compared results from various models,  both for the interatomic forces and for the electronic hopping parameters, using the same twist angle, $\theta = 1.05^\circ$. This choice is motivated by the fact that the value of the twist angle is the magnitude most accessible experimentally. It is yet unclear how precisely the experimentally studied twist angles correspond to the theoretical definition of magic angles. The dependence found here of the electronic properties on the choice of parameters suggests that the observed tendency towards broken symmetry phases must be quite robust. The appearance of superconductivity and insulating behavior in twisted graphene bilayers is likely to arise from rather general properties of the models.

\acknowledgments
FG  was  supported  by
the European Commission under the Graphene Flagship,
contract CNECTICT-604391; NRW is supported by  UK STFC under grant  ST/P004423/1.
\appendix
\section{Analytical model for lattice deformation}\label{sec:NK}
Here we derive an analytic expression for the elastic deformation of bilayer graphene, based on the work by Nam and Koshino \cite{nam_lattice_2017}.

We assume that the lattice vectors of the two unperturbed graphene lattices, which are rotated by a relative angle $\theta$, for each layer are given by (from  now we use  the graphene lattice spacing, $1.42\sqrt{3}\,\text{\AA}$, as a length unit) 

\begin{equation}
\vec a_1=R_{-\theta/2}(1,0),\vec a_2=R_{-\theta/2}(1/2,\sqrt{3}/2).
\end{equation}
and for the second layer we have
\begin{equation}
\tilde{\vec a}_i=R_{\theta} \vec a_i.
\end{equation}
The lattice vectors of the super cell are
\begin{equation}
\vec b_1=m \vec a_1 + n\vec a_2, \vec b_2 =(n+m) \vec a_1 -m  \vec a_2,
\end{equation}
and the angle between the two layers can be expressed as
\begin{equation}
\theta=\cos^{-1}\left(\frac{m^2 + n^2 + 4 m n}{2 m^2 + 2 n^2 + 2 m n}\right).
\end{equation}
We can also express this in terms of $\tilde{\vec a}_i$, where $m$ and $n$ change roles:
\begin{equation}
\vec b_1=n \tilde{\vec a}_1 + m\tilde{\vec a}_2, \vec b_2 =(m + n)\tilde{\vec a}_1 - n \tilde{\vec a}_2,
\end{equation}
In the remainder we shall always implicitly assume that the angle $\theta$ is small (normally we will only consider the case $m=n+1$ where $\theta=\sin^{-1}\left(\frac{\sqrt{3}(2m-1)}{6 m^2-6 m+2}\right)\approx1/(\sqrt{3}m)$). We will denote $R_{\theta/2}$ as $R$.

\begin{figure}
\includegraphics[width=6cm]{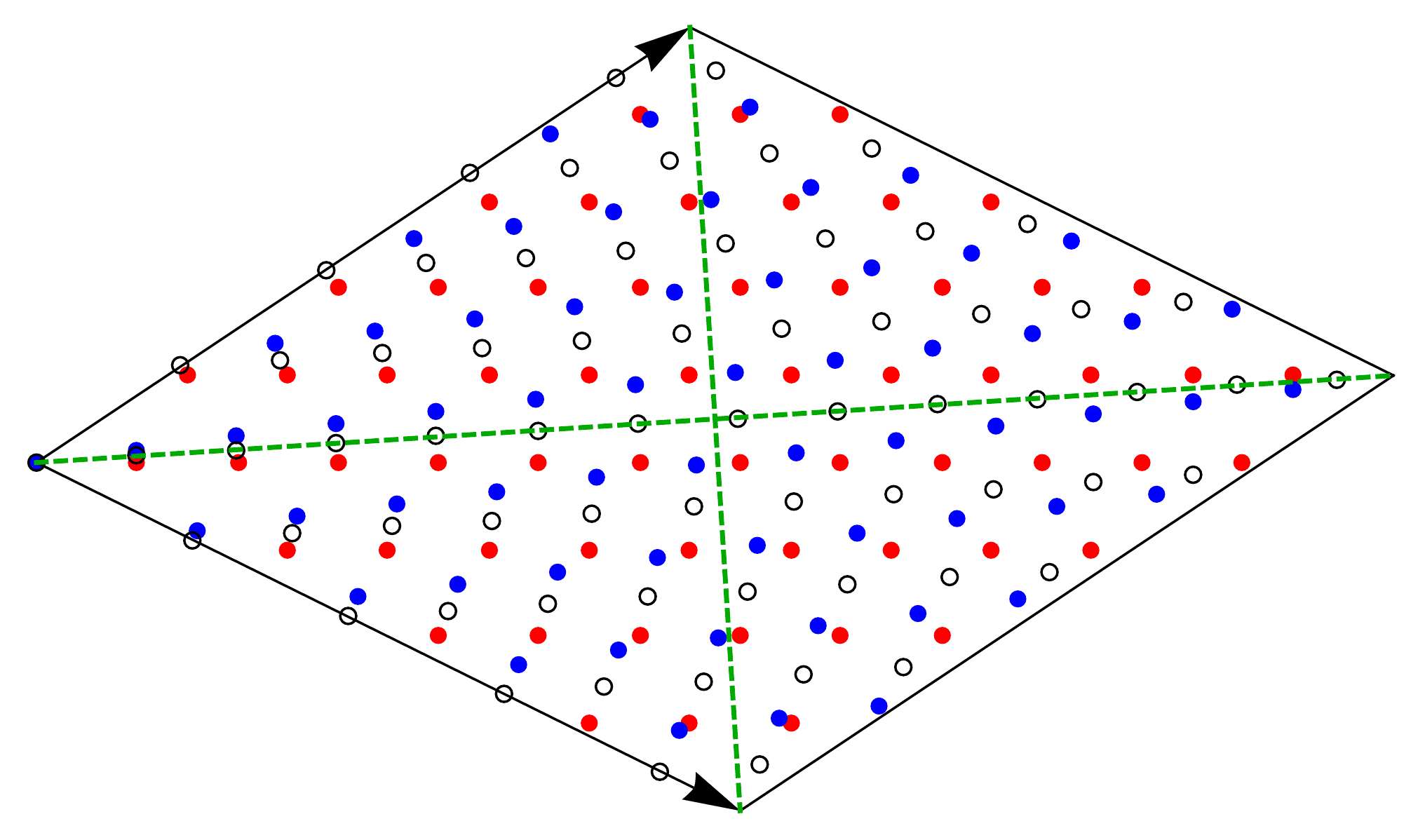}
\caption{An example of a supercell in an $(m,n)=(5,4)$ grid. Red points show the $+$ lattice and blue points the $-$ one. The circles are the average positions. Note the reflection symmetries in the two green lines, which are broken for the average positions.}\label{fig:symm}
\end{figure}

There is a symmetry between the layers, as can be seen in Fig.~\ref{fig:symm}. We label  the layers by $+$ (top) and $-$ (bottom). It is easy to show that with the lattice positions given by $\vec r^{(0)+} _{kl}=k \vec a_1 + l\vec a_2$ and $\vec r^{(0)-}_{kl}=k \tilde{\vec a}_1 + l\tilde{\vec a}_2$ we have an additional symmetry under reflection $T_x$ in the $x$-axis, 
\begin{equation}
T\vec r^{(0)+} _{kl}=\vec r^{(0)-} _{k,-l},
\end{equation}
We have a similar symmetry for reflections in the line connecting $\vec{b}_1$ to $\vec{b}_2$, Without writing down the detailed form of the transformation matrices, we see that this maps 
\begin{equation}
r^{(0)+}_{kl}\rightarrow r^{(0)-}_{-k+3(m+n)/2,l+3(m-n)/2}.
\end{equation}

We now assume that the two lattices will deform in a similar way, respecting the reflection symmetry. If we define an average lattice by the vectors 
\begin{equation}
\bar{\vec a}_i=(\vec a_i+\tilde{\vec a}_i)/2=(R^{-1}+R)/\tilde{\vec a}_i/2=
\cos(\theta/2) R^{1/2} \vec a_i,
\end{equation}
We can write for the lattice vectors in the two lattices, labeled as $\pm$,
\begin{equation} 
\vec r^\pm _{kl}=\vec{r}^{(0)\pm}_{kl} +\vec u^\pm(\vec{r}^{(0)\pm}_{kl})=\vec r^{(0)}_{kl} \pm \frac{1}{2}\delta r^{(0)}_{kl}+\vec u^\pm(\vec{r}^{(0)\pm}_{kl}),
\end{equation}
with $\vec r_0=k\bar{\vec a}_1+l\bar{\vec a}_2$. If we assume $\delta r^{(0)}_{kl}$ is small\footnote{That is not true over the whole supercell; the misalignment is one lattice spacing at the far corner of the supercell. Fortunately, that is where $r$ is large as well.}, then we can make the approximation that $\vec u^\pm(\vec{r}^{(0)\pm}_{kl})=\pm \vec u(\vec{r}^{(0)}_{kl})$, and we can simplify this expression. We use $kl$ as labels to show that their range is either $(m,n)$  or $(n,m)$, depending on the layer. As we can see from Fig.~\ref{fig:symm}, this makes most sense in half the Brillouin zone; we can, however, work with the hexagonal Brillouin zone where this approach works well everywhere.

We define the three reciprocal lattice vectors
$
\vec g_i
$
to $\vec{a}_j$,  and similar for $\tilde{\vec{a}}_j$.
We then define the superlattice reciprocals,
\begin{equation}
\vec{G}_i=\frac{1}{m-n}(1-R(\theta))\vec g_i
\end{equation}
It is straightforward to see that $\vec{G}_i\cdot (\vec g_i+\tilde{\vec{g}}_i)=0$.
[Note the slightly awkward labeling: $\vec{G}_1$ and $\vec{G}_3$ are the dual vectors to $\vec{b}_1$ and $\vec{b}_2$.]

We now minimize the combination of the misalignment of the lattices and the elastic energy as done by Nam and Koshino, with a minor change in the vectors used in the misalignment energy, assuming that we can write the continuum approximation (notice that here there is an important difference with Koshino, who have no reference to the mean displacements, but work in one of the two sub-lattices, so the meaning of $\vec{r}$ is very different, and their final results lacks the layer symmetry found below)
\begin{equation}
\vec \delta(\vec r)=\vec\delta_0(\vec r)+(\vec u^+-\vec u^-)(\vec r),
\end{equation}
where $\vec \delta(\vec r)$ is a field in the average lattice, with $\vec \delta(\vec r) \cdot\vec r=0$.
Since the $\vec \delta$ is the vector from the top to the bottom lattice, we would like to align this displacement with the favourable positions for the 
top lattice, but then we would like to align $-\vec\delta$ with the bottom lattice. Thus we see we need to minimise the potential 
\begin{align}
V[\delta]&=V_0 \sum_{j=1}^3 (\cos({\vec{g}}_j\cdot \vec \delta)+\cos(\tilde{\vec{g}}_j\cdot \vec \delta))\nonumber\\
&=V_0 \sum_{j=1}^3 2\cos\left(({\vec{g}}_j+\tilde{\vec{g}}_j)/2\cdot \vec \delta\right)\cos\left(({\vec{g}}_j-\tilde{\vec{g}}_j)/2\cdot \vec \delta\right)\nonumber\\
&\approx 2V_0 \sum_{j=1}^3 \cos\left(({\vec{g}}_j+\tilde{\vec{g}}_j)/2\cdot \vec \delta\right)
\end{align}
We find that, using the average $\bar{\vec g}_j=(\vec g_j+\tilde{\vec g}_j)/2$, 
 \begin{eqnarray}
\bar{\vec{g}}_j\cdot\vec \delta_0(\vec r)&=& -\frac{1}{2} ((I+R )\vec g_j)\cdot ((I-R)(k \vec a_1+l \vec a_2),\\
&=&((I-R)\vec g_j)\cdot ((I+R)(k \vec a_1+l \vec a_2)/2\nonumber\\
&=&
\vec{G}_j \cdot \vec r\,.
\end{eqnarray}
Thus,
\begin{equation}
V[\delta]= 2V_0\sum_{j=1}^3 \cos\left({\vec{G}}_j\cdot \vec r +\bar{\vec g}_j \cdot \vec u(\vec r)\right),
\end{equation}
We can now follow Nam and Koshino, and the standard continuum elastic energy to the energy derived here. This lead to the requirement to solve the coupled equations, where $\vec{q}_\perp=(q_y,-q_x)$:
\begin{align}
\sin(\vec G_j\cdot \vec r+\bar{\vec g}_j\cdot \vec u(\vec r))&=\sum _{\vec q} f_{\vec q}^j e^{i\vec q \cdot \vec r},\\
\vec u(\vec r)&=\sum_{\vec q} \vec u_{\vec{q}} e^{i\vec q\cdot\vec r},\label{eq:deff}\\
\vec{u}_{\vec{q}}=4V_0\sum_{j=1}^3 f^j_{\vec q} \frac{1}{q^4} 
&\left[\frac{1}{\lambda+2\mu} \vec q \vec q^T\bar{\vec g}_j
+\frac{1}{\mu}{\vec q}_\perp {\vec q}_\perp^T \bar{\vec g}_j\right].
\end{align}
If we make the simplest approximation for the sine, neglecting completely the contribution from $\vec u$, we find that
\begin{equation}
f^j_{\vec q} =\delta_{\vec q, \pm \vec G_j}\frac{\pm1}{2i}, 
\end{equation} 
and thus, since $\vec{G}_j$ and $\bar{\vec g}_j$ are orthogonal, we find that ($(1)$ for first order)
\begin{equation}
\vec u^{(1)}_{\vec q}=\frac{4V_0}{\mu}\delta_{\vec q, \pm \vec G_j}\frac{\pm1}{2i}\frac{1}{G^4}(G^j_y,-G^j_x)(G^j_y,-G^j_x)\cdot \bar{\vec g}_j
\end{equation}
Since the two vectors $(G^j_y,-G^j_x)$ and $\bar{\vec g}_j$ are parallel, this can be written as
\begin{align}
\vec u^{(1)}_{\vec q}&=\frac{4V_0}{\mu}\delta_{\vec q, \pm \vec G_j}\frac{\pm1}{2i}\frac{1}{G^4\bar g^2}\left[(G^j_y,-G^j_x)\cdot \bar{\vec g}_j\right]^2
\bar{\vec g}_j\nonumber\\
&=\frac{4V_0g}{\mu G^2}\delta_{\vec q, \pm \vec G_j}\frac{\pm1}{2i}
\hat{\bar{\vec g}}_j.
\end{align}
Thus we find that the dimensionless quantity\begin{equation}
g \vec u^{(1)}(\vec r)=\frac{4V_0g^2}{\mu G^2} \sum_j^3\hat{\bar{\vec g}}_j \sin( \vec G_j\cdot\vec r).
\end{equation}
The expansion parameter $\alpha=\frac{4V_0g^2}{\mu G^2}$ grows with the size of the unit cell, showing that for very small angles a perturbative approach must fail.

\begin{figure}
\includegraphics[width=5cm]{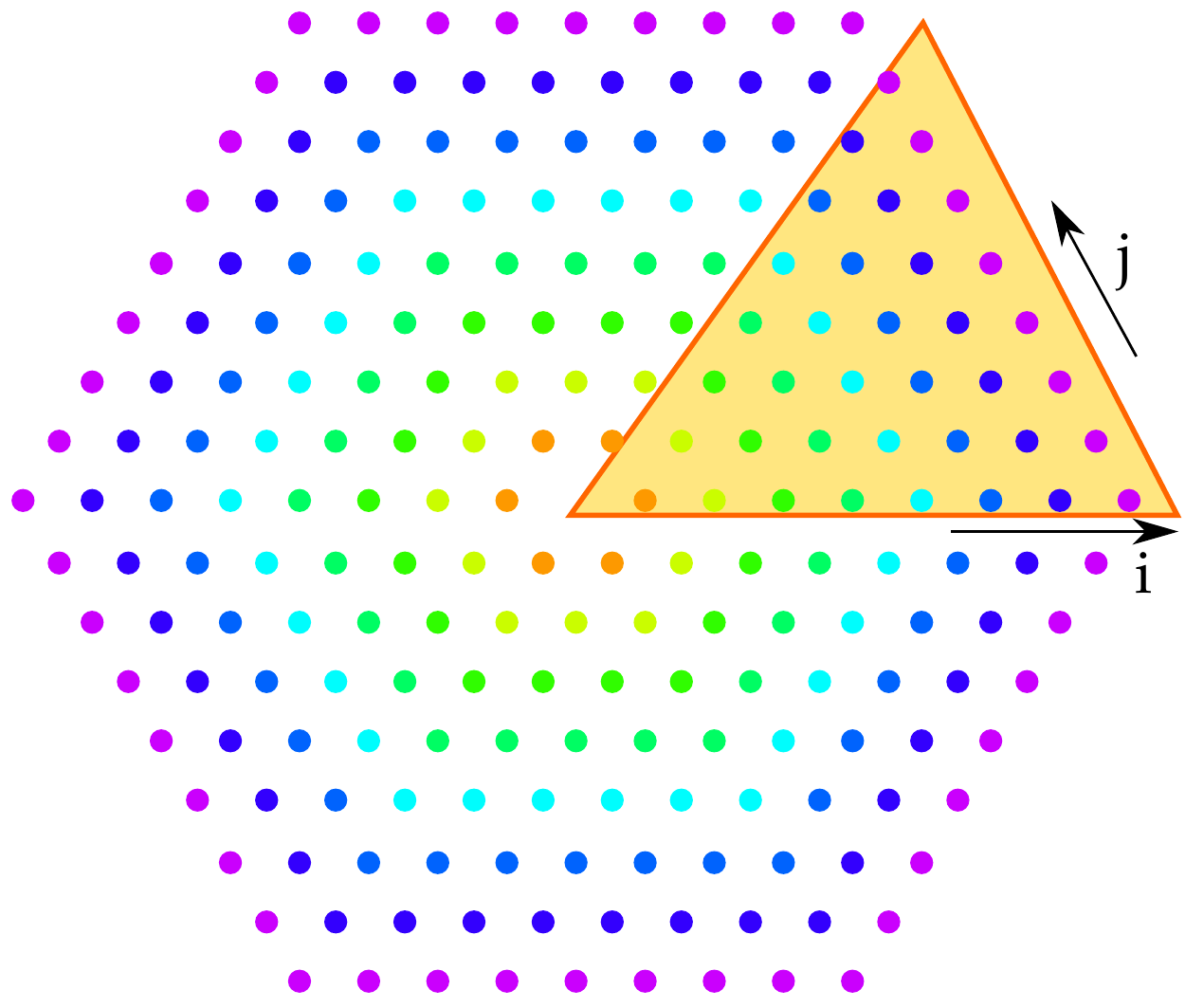}
\caption{points $q$ used in Table~\ref{tab:uq}}\label{fig:kpoints}
\end{figure}

With the help of a simple mathematica code it is now straightforward to find the higher order terms, which involves expanding Eq.~\eqref{eq:deff} to higher order in $V_0/\mu$.
Results following the notation by Nam and Koshino are given in Table.~\ref{tab:uq}. Our results are a universal (lattice-size independent) expression when we  scale $u_{\vec q}$ as $\frac{g}{i}u_{\vec q}$, and express the values in terms of the parameters
\begin{eqnarray}
\alpha&=&4\frac{V_0}{\mu} g^2/G^2=\frac{V_0}{\mu}\cot^2(\theta/2),\\
\beta&=&\mu/(\lambda+2\mu).\end{eqnarray}
Here we use $
g^2=\frac{8\pi^2}{3a^2}(1+\cos\theta)$ and $
G^2=\frac{64\pi^2}{3a^2}\sin^2\theta/2$.
\newcommand{\ifrac}[2]{(#1/#2)}
\begin{table*}
\caption{The results for $\frac{g}{i}u_{\vec q}$ using fourth order perturbation theory for $q$ vectors in one sixth of space.}
\label{tab:uq}
\begin{ruledtabular}
\begin{tabular}{cc}
	(1,0) & $-\ifrac{\alpha}{128}   \left(31 \alpha ^3-36 \alpha ^2-16 \alpha +64\right) (0,1) $\\
	(2,0) & $\ifrac{\alpha ^2 \left((3 \beta+1579) \alpha ^2+392 \alpha -1568\right)}{25088} (0,1) $\\
	(2,1) & $\ifrac{\alpha ^2 \left((3 \beta-1277) \alpha ^2-1176 \alpha +1568\right)}{75264} \left(\sqrt{3},-3\right) $\\
	(3,0) & $-\ifrac{\alpha ^3 (\alpha  (3 \beta-59)+392)}{37632} (0,1) $\\
	(3,1) & $\ifrac{\alpha ^3}{2458624} \left(\sqrt{3} \left(392 (19-5 \beta)+\alpha  \left(135 \beta^2+276 \beta-6193\right)\right),\alpha  \left(81 \beta^2+216 \beta+30965\right)-392 (3 \beta+95)\right) $\\
	(3,2) & $\ifrac{\alpha ^3}{1229312} \left(\sqrt{3} \left(392 (19+2 \beta)+\alpha  \left(-54 \beta^2-123 \beta-6193\right)\right),\alpha  \left(-81 \beta^2-153 \beta+12386\right)+392 (3 \beta-38)\right) $\\
	(4,0) & $-\ifrac{1}{512} \alpha ^4 (0,1) $\\
	(4,1) & $\ifrac{\alpha ^4}{8479744} \left(7 \sqrt{3} (655-3 \beta (\beta+96)),-3 \beta (3 \beta+340)-32095\right) $\\
	(4,2) & $\ifrac{\alpha ^4 (3 \beta-185)}{150528} \left(-\sqrt{3},3\right) $\\
	(4,3) & $\ifrac{\alpha ^4}{8479744} \left(3 \sqrt{3} (\beta (5 \beta+506)+4585),3 \beta (9 \beta+838)-22925\right) $\\
\end{tabular}
\end{ruledtabular}
\end{table*}

When using this for finite discrete lattices, we shall use $\vec r^{(0)\pm}$ as the argument of $\vec u$, which restores the broken reflection symmetry.
\bibliography{GrapheneDeformation}
\end{document}